\documentclass[%
 reprint,
 amsmath,amssymb,
 aps,
floatfix,
]{revtex4-2}

\usepackage{graphicx}
\usepackage{dcolumn}
\usepackage{bm}
\usepackage{amsmath}
\usepackage{amsfonts}
\usepackage{subfiles}
\usepackage{xcolor}
\usepackage{dsfont}
\usepackage[utf8]{inputenc}
\usepackage{dirtytalk}
\usepackage{ulem}
\usepackage{comment}
\newcommand{\bs}[1]{\boldsymbol{#1}}
\newcommand{\ev}[1]{\left\langle#1 \right\rangle}

\newcommand{\targ}{\text{targ}}
\newcommand{\out}{\text{out}}
\usepackage{diagbox}
\usepackage{enumitem}
\usepackage{hyperref}

\newcommand{\edit}[1]{{\color{black}{#1}}}
\newcommand{\editt}[1]{{\color{black}{#1}}}
\begin{document}
\preprint{APS/123-QED}
\title{Quantum reservoir computing using arrays of Rydberg atoms}

\author{Rodrigo Araiza Bravo$^{1}$}
 \email{oaraizabravo@g.harvard.edu}
\author{Khadijeh Najafi$^{1,2}$}
\email{knajafi@ibm.com}
\author{Xun Gao$^{1}$}
\author{Susanne F. Yelin$^{1}$}

\affiliation{%
$^{1}$Department of Physics, Harvard University, Cambridge, Massachusetts 02138, USA\\
$^{2}$IBM Quantum, IBM T.J. Watson Research Center, Yorktown Heights, NY 10598 USA}%
\date{\today}

\begin{abstract}
Quantum computing promises to speed up machine learning algorithms. However, noisy intermediate-scale quantum (NISQ) devices pose engineering challenges to realizing quantum machine learning (QML) advantages. Recently, a series of QML computational models inspired by the noise-tolerant dynamics of the brain has emerged as a means to circumvent the hardware limitations of NISQ devices. In this article, we introduce a quantum version of a recurrent neural network (RNN), a well-known model for neural circuits in the brain. Our quantum RNN (qRNN) makes use of the natural Hamiltonian dynamics of an ensemble of interacting spin-1/2 particles as a means for computation. In the limit where the Hamiltonian is diagonal, the qRNN recovers the dynamics of the classical version. Beyond this limit, we observe that the quantum dynamics of the qRNN provide it with quantum computational features that can aid it in computation. To this end, we study a \editt{fixed geometry qRNN, i.e. a quantum reservoir compute, based on arrays of Rydberg atoms  and show that the Rydberg reservoir} is indeed capable of replicating the learning of several cognitive tasks such as multitasking, decision-making, and long-term memory by taking advantage of several key features of this platform such as interatomic species interactions, and quantum many-body scars.
\end{abstract}
\maketitle

\section{Introduction}
Quantum computing promises to enhance machine learning algorithms. However, implementing these advantages often relies on either fault-tolerant quantum computers not yet available \cite{biamonte2017quantum,Harrow2009,Wiebe2012,Low2014,Lloyd2014}, or on decoherence-limited, variational quantum circuits which may experience training bottlenecks \cite{mcclean2018barren,cerezo2021cost}. Thus, currently available noisy intermediate-scale quantum (NISQ) devices thwart quantum advantages in machine learning algorithms. \par
Recently, to counteract these challenges, several quantum machine learning architectures have emerged inspired by models for computation in the brain \cite{Markovic2020A, Kiraly2021, mujal2021opportunities}. These brain-inspired algorithms are motivated by the inherent robustness of input- and hardware-noise in brain-like computation, and by the possibility to use the analogue dynamics of controllable, many-body quantum systems for computation without relaying on a digital circuit architecture. Broadly speaking, these brain-inspired algorithms can be put into two categories. The first of which encompasses systems quantizing the dynamics of biological computational models at the single-neuron level. Thus, the dynamics of single qubits or groups of qubits resemble the dynamics of a neurons in a neural circuit of interest. Examples of these include quantum memristors \cite{pfeiffer2016quantum}, which are electrical circuits with a history-dependent resistance, quantum versions of the biologically realistic Hodgkin-Huxley model for single neurons \cite{gonzalez2019quantized,gonzalez2020quantized}, and unitary adiabatic quantum perceptron \cite{torrontegui2019unitary}. \par 
The second category of brain-inspired algorithms relies on a macroscopic resemblance between many-body quantum systems and neural circuits. In this regard, the algorithms that have received the most attention are quantum reservoir computers. Quantum reservoir computers use ensembles of quantum emitters with fixed interactions to perform versatile machine learning tasks relying on the complexity of the unitary evolution of the system. Since these systems can couple with both classical and quantum devices, which may encode the tasks' input, quantum reservoirs have been used for time-series prediction \cite{fujii2020quantum, Nakajima2019,Kutvonen2020}, entanglement measurement \cite{Ghosh2019A, khan2021physical}, quantum state preparation \cite{Gosh2019B}, continuous-variable computation \cite{govia2021quantum} which can be made universal \cite{nokkala2021gaussian}, reduction of depths in quantum circuit \cite{Gosh2020}, ground state finding \cite{mujal2022quantum}, and for long-term memory employing ergodicity-breaking dynamics \cite{sakurai2021quantum,xia2022reservoir,PhysRevLett.127.100502}. See \cite{mujal2021opportunities} for a comprehensive review of quantum reservoir computing. \par
In both categories, however, a thorough understanding of the potential computational advantages and their origins are slowly emerging. In this article, we contribute to this direction by proposing a quantum extension of a well-known neural circuit model called recurrent neural networks (RNNs), of which reservoir computers are a special case \cite{coolen2001}. Our extension uses the Hamiltonian dynamics of ensembles of two-level systems. In the limit where the Hamiltonian is diagonal, we recover the classical single-neuron dynamics \edit{ naturally encoding RNNs into quantum hardware. Recently, another natural encoding of a reservoir computer was proposed using superconducting qubits \cite{suzuki2022natural}. In our case,} the general dynamics of the quantum RNN (qRNN) present several new features that can aid in the computation of both classical and quantum tasks. In particular, a qRNN used for simulating stochastic dynamics can exhibit speedups compared to classical RNNs. \par 
To show that our scheme is experimentally realizable, we propose that arrays of Rydberg atoms can be used as qRNNs (Sec. \ref{sec:Rydberg}). \edit{Although our Rydberg qRNNs have restricted connectivity, we are motivated to use Rydberg arrays due to recent studies with equally restricted qRNNs which show significant computational capacity when driven near criticality \cite{Kutvonen2020, Kutvonen2020, mujal2022quantum}.} Moreover, recent experiments using optical tweezers \cite{Saffman2010, lester2015rapid, barredo2016atom,endres2016atom,labuhn2016tunable, Bernien2017,cooper2018alkaline, wilson2022trapping} have catapulted the community’s interest in Rydberg arrays as they exhibit long coherence times, controllable and scalable geometries, and increasing levels of single-atom control \cite{Ebadi2020}. Additionally, Rydberg arrays can be used for a novel, programmable quantum simulations and universal computations \cite{isenhower2010demonstration, Saffman2010, Pichler2018,Omran2019, henriet2020quantum, cohen2021quantum}. \par
We numerically \editt{implement fixed-geometry Rydberg qRNNs, i.e. Rydberg reservoir computers,} and we successfully perform cognitive tasks even when a few atoms are available (Sec. \ref{sec:BioTasks}). 
The success of these tasks is explained by the physics of Rydberg atoms. For example, our Rydberg qRNNs excel at learning to multitask since they can naturally encode RNNs with inhibitory and excitatory neurons which are vitals for many cognitive tasks \cite{Song2016_ExInh}. This encoding relies on the different types of interactions between Rydberg atoms with different principal quantum numbers \cite{Han2009}. Likewise, a Rydberg qRNN exhibits long-term memory due to the weak-ergodicity breaking dynamics of many-body quantum scars \cite{Bernien2017,bluvstein2021controlling, maskara2021discrete}. Lastly, we discuss possible further research directions in Sec. \ref{sec:Conclusion}. \par
We remark that the notion of qRNNs has been previously coined relying on universal quantum circuits and using measurements to implement the nonlinear dynamics of an RNN \cite{bausch2020recurrent}. Instead, what we define as a ``quantum RNN'' leverages the inherent unitary dynamics of ensembles of two-level systems to compute, deviating from the quantum digital circuit model for computation. \par 

\section{Classical recurrent neural networks}\label{sec:ClassicalRNN}
We begin by reviewing an archetypal RNN consisting of $N$ binary neurons. Each neuron is in one of two possible states $s_n(t)\in\{-1,1\}$ and is updated from the time-step $t$ to $t+1$ following the update rule 
\begin{align}\label{eq:cupdate}
    s_n(t+1)&= \text{sign}\left(h_n(t)s_n(t)\right),\notag\\
    h_n(t)&\equiv -\Delta_n(t)+\sum_{m}J_{nm}s_m(t),
\end{align}
where $J_{nm}=J_{mn}$ are symmetric synaptic connections between neurons $n$ and $m$. The time-dependent biases $\Delta_n(t)$ encode the RNN’s inputs. To avoid memorization during a learning task with inputs $u_n^{\text{task}}(t)$, the RNN receives Gaussian-whitened inputs 
\begin{equation}\label{eq:inputs}
\Delta_n(t) = u_n^{task}(t)+\xi_n,
\end{equation}
where \edit{$\xi_n$ is a zero-mean Gaussian random variable with variance $\sigma_{in}^2$, making the evolution of the RNN stochastic. In RNNs, the value of $\sigma_{in}^2$ is proportional to the value of the tasks' inputs $u_n^{task}$.} \par
When studying learning tasks similar to those in the mammalian cortex \cite{Song2016_ExInh} one turns to a continuous version of the rule in (\ref{eq:cupdate}) obtained in the case that the time-interval $\tau$ in which neurons update is small \edit{compared to $J_{nm}$}. In this limit,
\begin{equation}\label{eq:EOMClassical}
    \tau \dot{s}_n(t)= -s_n(t)+\text{sign}\left(h_n(t)s_n(t)\right).
\end{equation}
Thus, the RNN obeys a system of nonlinear differential equations. Note that (\ref{eq:EOMClassical}) imply that $s_n\in[-1,1]$ is a continuous and bounded variable \cite{coolen2001}. \par 
A third way to describe an RNN is via the probability distribution $p_{t}(\bs{s})$ of observing each of the $2^N$ different configurations $\bs{s}$ at the $t^{th}$ time-step. \edit{Due to the noise in the inputs $\Delta_n$, the dynamics of the distribution follow a Markov chain description \cite{coolen2001}}. \edit{This description is particularly useful for analyzing the stochastic dynamics simulatable by an RNN}. As we shall see in Sec \ref{sec:QAstochastic}, this representation will be useful in explaining how, relative to classical RNNs, the unitary dynamics of a qRNN can speed up \edit{stochastic process simulations.}\par 
Lastly, we describe how to use an RNN for computation. After the RNN evolves for a time $t_f$, a subset of $M$ neurons are used to collect the vector \edit{$\bs{r}(t_f) = (s_{n_1}(t_f),...,s_{n_M}(t_f), 1)$ with the last entry accommodating for a bias.} The other $N-M$ other neurons are called \textit{hidden neurons}. The RNN's output is obtained via a linear transformation $\bs{y}^{\out} = W^{\out}\bs{r}(t_f)$ where $W^{\out}$ is a real-valued matrix. Thus, the computational complexity of the RNN comes from the nonlinear activation function in (\ref{eq:cupdate}) which enables $\bs{y}^{\text{\out}}$ to be a nonlinear function of the inputs.\par 
In a learning task with a target output $\bs{y}^{\targ}$, the RNN is trained by minimizing a loss function $\mathcal{L}(\bs{y}^{\text{out}}, \bs{y}^{\text{targ}})$ with respect to the network parameters such as $W^{\text{out}}$, $J_{nm}$, etc.  \edit{subject to the task-determined inputs in (\ref{eq:inputs})}. We choose the square-loss 
\begin{equation}\label{eq:Loss}
\mathcal{L}(\bs{y}^{\out}, \bs{y}^{\text{targ}}) = \frac{1}{N_s}\sum_{i=1}^{N_{s}}||\bs{y}_i^{\text{targ}}-\bs{y}_i^{\out}||^2,
\end{equation}
where $i$ labels the $N_s$ different input instances. \edit{For the tasks in Sec. \ref{sec:BioTasks}, we fix the connections $J_{nm}$, such that our qRNNs more closely resemble quantum reservoir computers.}  \par 
\section{Quantum recurrent neural networks}\label{sec:QuantumRNN}
\subsection{Quantum update rule}\label{sec:QUpdate}
Let us now extend the classical RNN in (\ref{eq:cupdate}) to the quantum setting. We replace each of the $N$ neurons with a spin-1/2 particle for which a spin measurement along the $z$-axis yields the values $\{-1,1\}$. Thus, each neuron $n$ is in a normalized quantum state in the Hilbert space $\mathcal{H}_n$ with basis vectors $\{|\text{-}1\rangle_n,|1\rangle_n\}$ which are eigenstates of the Pauli-Z operator $\sigma_n^z=|1\rangle\langle 1|_n-|\text{-}1\rangle\langle \text{-}1|_n$. The state of the composite system lives in the product Hilbert space $\mathcal{H}=\bigotimes_{n=1}^N \mathcal{H}_n$. \par 
We choose spins interacting via the time-dependent Hamiltonian 
\begin{align}\label{eq:HGeneral}
    H(t) &=-\sum_{n=1}^N\Delta_n(t)\sigma_n^z+\sum_{nm}J_{nm}\sigma_n^z\sigma_m^z \notag\\
    &+\frac{\Omega(t)}{2}\sum_{n=1}^N \sigma_n^x,
\end{align}
where $\sigma_n^x=|1\rangle\langle\text{-}1|_n+|\text{-}1\rangle\langle 1|_n$ is the Pauli-X operator. Indeed, the evolution under (\ref{eq:HGeneral}) encompasses the update rule in (\ref{eq:cupdate}). To see this, note that in the classical case of (\ref{eq:cupdate}), the RNN evolves under the rules
\begin{align}
 &\text{If } h_n>0,\text{ } s_n \text{ doesn't change.}         \tag{1C}\label{eq:1C} \\
 &\text{If } h_n<0,\text{ } s_n \text{ flips.}                  \tag{2C}\label{eq:2C}
\end{align}
Here ``$C$'' stands for ``classical''. Now, consider a qRNN starting in the configuration $|s_1,s_2,...,s_N\rangle$ and evolving for a time $t = 2\pi\Omega^{-1}$. In the limit where \edit{$\Delta_n \gg \Omega$ or $J_{nm}\gg \Omega$}, each spin experiences the Ham7iltonian $H_n=h_n\sigma_n^z+\frac{\Omega}{2}\sigma_n^x$ where $h_n=-\Delta_n+\sum_m J_{nm}s_m$ is the effective field generated by the rest of the spins \edit{where $s_m$ stands for the measurement result of $\sigma_m^z$ on the initial configuration}. We then obtain the quantum update rules
\begin{align}
&\text{If } \edit{|h_n|}\gg \Omega, \text{ } \edit{|s_n\rangle} \text{ doesn't change.} \tag{1Q}\label{eq:1Q} \\
&\text{If } \edit{|h_n|}\ll \Omega, \text{ } \edit{|s_n\rangle} \text{ flips.}          \tag{2Q}\label{eq:2Q}
\end{align}
Here, ``Q'' stands for ``quantum''. Therefore, (\ref{eq:HGeneral}) can implement (\ref{eq:1C})-(\ref{eq:2C}) but without the use of the nonlinear activation function in (\ref{eq:cupdate}). Nonetheless, (\ref{eq:HGeneral}) allows for more general dynamics beyond the perturbative limit for which (\ref{eq:1Q})-(\ref{eq:2Q})  holds. We now highlight three features arising from the quantum evolution of the qRNN: (i) the ability to compute complex functions on the input by using quantum interference, (ii) exploiting the choice of measurement basis, and (iii) efficiently achieving stochastic processes inaccessible to classical RNNs with no hidden neurons.

\begin{figure}
\centering
\includegraphics[width=0.90\linewidth]{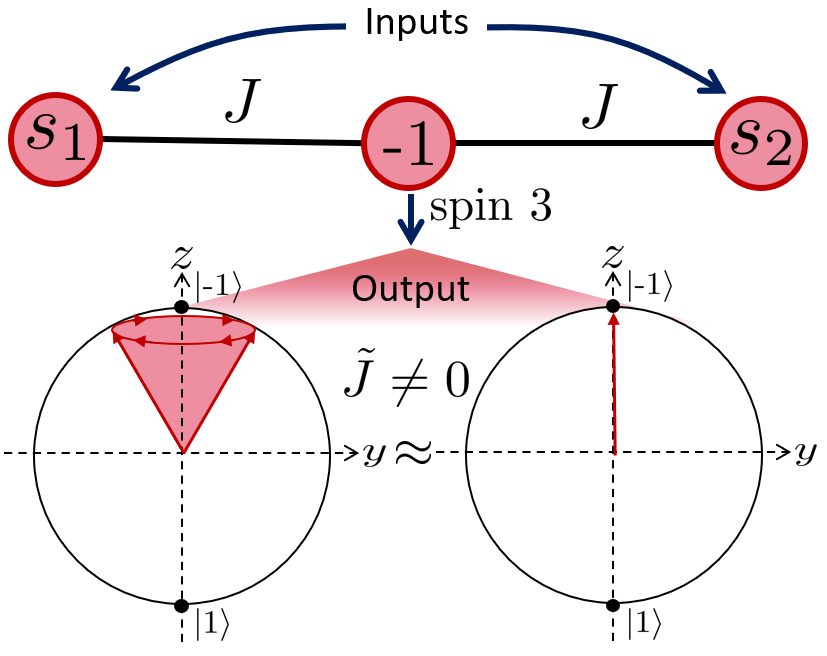}
\caption{Computing the parity, XOR($s_1,s_2$), of two inputs $s_1$ and $s_2$ with a qRNN. Spin 3 (the output spin) experiences an effective field $\tilde{J}= J(s_1+s_2)$ with $J\gg \Omega$. After evolving for a time $t=2\pi\Omega^{-1}$, we measure the output spin. The measurement outcome $\text{+}1$ is obtained when $s_1=-s_2$ since $\tilde{J}=0$. If $s_1=s_2$ so that $\tilde{J}\neq 0$, the inputs constructively interfere to generate a large detuning on the output such that measurement yields the outcome -1. 
}
\label{fig:XOR_classification}
\end{figure}
\subsubsection*{Quantum feature 1: quantum interference as a means for computation}\label{sec:QAbasis}
The computational power of (\ref{eq:cupdate}) is a result of its nonlinear dynamics. For example, an RNN with linear dynamics is incapable of computing the parity function $\text{XOR}(s_1,s_2)=s_1s_2$ between two classical binary inputs.  On the other hand, quantum mechanics is a unitary theory. Yet, this does not limit a qRNN to linear computation. Indeed, a qRNN can compute $\text{XOR}$ by leveraging quantum interference, a resource fundamental to quantum computation. Thus, we can use a qRNN for complex computing tasks. \par 
As illustrated in Fig. \ref{fig:XOR_classification}, we can compute XOR($s_1,s_2$) using a qRNN of three spins initially in the state $|s_1,s_2,\text{-}1\rangle$. The third spin is an outcome spin. This spin is measured to tell us information about the parity of $s_1$ and $s_2$. We let these spins evolve under the dynamics dictated by (\ref{eq:HGeneral}) choosing $\Delta_n, J_{12} = 0$ and $J_{13}=J_{23}=J\gg \Omega$. Let $\tilde{J}=J(s_1+s_2)$. In the frame rotating at the rate $\tilde{J}$, the output spin experiences the Hamiltonian
\begin{align}
    H_3 &=\frac{\Omega}{2}\left(e^{2i\tilde{J}\tau}|1\rangle\langle\text{-}1|+\text{h.c.}\right).
\end{align}
It's clear that if the spins have odd parity (i.e. $s_1=-s_2$ so that $\tilde{J}=0$), the output spin flips to the state $|1\rangle$ when we choose to evolve by $t=2\pi\Omega^{-1}$. On the other hand, if $\tilde{J}\neq 0$, $H_3$ contains only fast-rotating terms, and the rotating-wave approximation (RWA) allows us to neglect the evolution of the output spin \cite{wu2007strong}. Physically,  the RWA can be thought of as the spin rotating along the $x$-axis by a small amount followed by a rapid precession of the spin around the $z$-axis. Indeed, as illustrated in Fig. \ref{fig:XOR_classification}, $J\gg t^{-1}$ amounts to averaging out the spin's position so that the spin is along the $z$-axis. \edit{Overall, this computation realizes the operation $|s_1,s_2,-1\rangle\rightarrow |s_1,s_2,XOR(s_1,s_2)\rangle$}\par 
Note that this is a result of $s _{1}+s_2$ constructively interfering to produce a large effective detuning on the output and blocking its evolution. Thus, interference serves as a means for computation in qRNNs.

\begin{figure}[htp]
\centering
\includegraphics[width=0.90\linewidth]{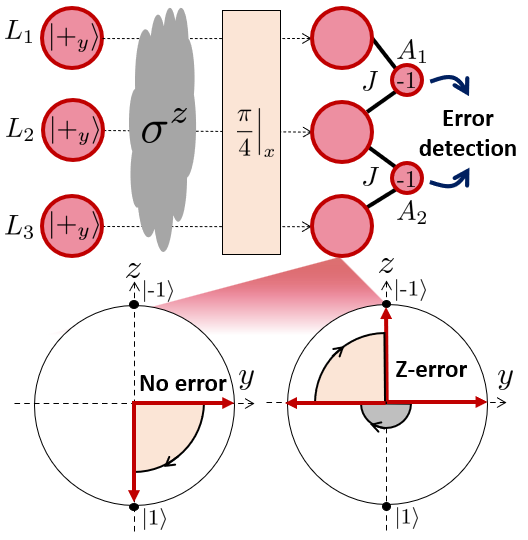}
\caption{\edit{Detection of a Z-error on three spins $L_{1,2,3}$ using a quantum RNN. A Z-error is conjugated into a bit-flip-like error using a Hamiltonian generating a rotation along the x-axis where $ t=\pi/2\Omega$ and $\Omega$ is the dominant field of the Hamiltonian. The state of each of the $L_{i}$ after the rotation (orange region) depends on whether a Z-error occurs, as it’s illustrated at the bottom of the figure. As exemplified here for $L_3$, a Z-error results in a spin flipping from what we would expect in the absence of errors. To detect the Z-error, a set of auxiliary qubits $A_{1,2}$ is brought in to perform a parity measurements of pairs $(L_1,L_2)$ and $(L_2,L_3)$. Since under no Z-error the parity measurements must match, the parity measurements allow us to detect the location of the Z-error as specified in Table \ref{tab:Errors}.}}
\label{fig:Error}
\end{figure}

\subsubsection*{Quantum feature 2: arbitrary measurement basis as a means for computation}

Equations (\ref{eq:1Q})-(\ref{eq:2Q}) recover (\ref{eq:cupdate}) when $t=2\pi\Omega^{-1}$. However, $t=2\pi\Omega^{-1}$ is not a necessary restriction. This freedom results in the ability to rotate each quantum neuron which can be used as means for \edit{computing on a different basis. Measuring on different bases reveals the quantum correlations enhancing the performance of a qRNN relative to its classical counterpart.} In this section, we show how to use the qRNN’s evolution to change the basis on which an error occurs. This freedom can detect a Z-error, an error proper to quantum computation. \par 
\edit{Consider the repetition code $|0_L\rangle=|\text{-}_y\rangle^{\otimes 3}$ and $|1_L\rangle=|+_y\rangle^{\otimes 3}$ on qubits labeled $L_{1,2,3}$ where $|\pm_y\rangle = \frac{1}{\sqrt{2}}(|\text{-}1\rangle\pm i|1\rangle)$. Suppose we prepare the state $|\psi\rangle = a|0_L\rangle+b|1_L\rangle$, and consequently a Z-error occurs, we can detect the error by rotating all three spins $L_{1,2,3}$ using (\ref{eq:HGeneral}) with the dominant field being $\Omega$ for a time $t=\pi/2\Omega$. Note that the rotation conjugates the Z-error by 
\begin{equation}\label{eq:conjugate}
    e^{-i\pi\sigma^x/4}\sigma^ze^{i\pi\sigma^x/4} \propto \sigma^y
\end{equation}
where $\sigma_n^y=i|\text{-}1\rangle\langle 1|-i|1\rangle\langle\text{-}1|$ is like a bit-flip error except for a state-dependent phase. A bit-flip error can then be detected by bringing two extra spins $A_{1,2}$ and performing parity measurements of the pairings $(L_1,L2)$, and $(L_2,L_3)$ as described in Sec. \ref{sec:QAbasis}. Using Table \ref{tab:Errors}, the final parity of $(L_1,L2)$, and $(L_2,L_3)$ gives the measurement results $a_1$ and $a_2$ which can be used to discern where the Z-error occurred.} \par 
As an example, Fig. \ref{fig:Error}  illustrates the two final states of $L_3$ if no error occurs (bottom left), and if a Z-error occurs on $L_3$ (bottom right). \par 
Detecting the Z-error hinges on (\ref{eq:conjugate}) can be achieved by using the qRNNs evolution to rotate the measurement basis. Note that rotation allows us to measure the error syndrome of the stabilizer state $|\psi\rangle$, bringing out the quantum correlations of the state. Thus, the qRNN’s native evolution can be used to perform quantum computational tasks.  \edit{After the error is detected on spin $L_i$, all qubits are rotated again by $U^\dagger$ and $\sigma_i^z$ can be applied to correct the error.}\edit{ We note that using a repetition code for error detection is a well-known technique in the quantum computing community.}\par
\edit{The previous two quantum features show that qRNNs are naturally suited to solve important problems in machine learning and quantum computing. Recently, qRNNs were used to compress quantum circuits \cite{Gosh2020}. However, studies on using qRNNs for error correction in circuit-like quantum computing are warranted and left for further studies.
}

\begin{table}[ht]
\centering
\begin{tabular}{|c||c|c|}
\hline
    \diagbox{$a_2$}{$a_1$} & -1 & +1 \\\hline\hline
     -1 & \edit{Error in $L_2$}         & Error in $L_1$   \\\hline
     +1 & Error in $L_3$   & \edit{No error}   \\
\hline
\end{tabular}
\caption{Results of parity measurements for detection of a Z-error. Measuring spin $A_i$ results in the outcome $a_i$. By comparing the outcomes, one can detect the location of the Z-error.}
\label{tab:Errors}
\end{table}

\begin{figure}[htp]
\centering
\includegraphics[width=0.90\linewidth]{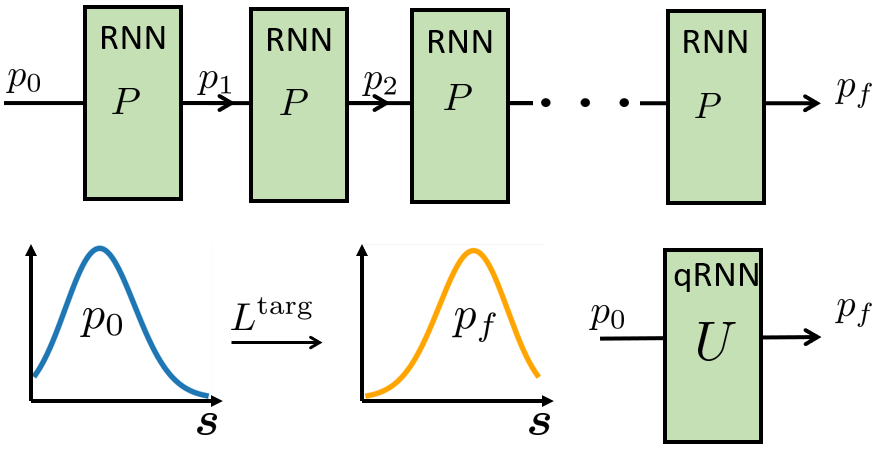}
\caption{Comparing a classical and a quantum RNN to \edit{stochastically evolve} a distribution $p_{t_f}$ from an initial distribution $p_0$. Here, we consider $p_{0}(\bs{s})=p_{t_f}(\bs{s}^\prime)$ when $s_n = -s^\prime_n$ for all $n$. In this case, the RNN needs to produce a stochastic process matrix $L^{\targ}$ that flips all the spins through several time-steps.  The classical RNN (top) requires $\mathcal{O}(2^{N-m})$ time-steps (i.e applications of $P$) while using $m$ hidden neurons. A qRNN (bottom) requires one time-step and no hidden neurons.}
\label{fig:training}
\end{figure}

\subsubsection*{Quantum feature 3: stochastic processes accessible to a qRNN}\label{sec:QAstochastic}
We now explore how a qRNN can be used to \edit{stochastically evolve} a probability distribution faster than any classical RNN. Firstly, we note that if we initialize an RNN according to an initial distribution $p_0(\bs{s})$, the dynamics in (\ref{eq:cupdate}) dictate that for $t>0$ the RNN obeys a distribution given by the Markov-chain dynamics 

\begin{align}\label{eq:Markov}
p_{t}(\bs{s}) = \sum_{s^{\prime}}P({\bs{s}|\bs{s}^{\prime}})p_{t-1}(\bs{s}^{\prime})
\end{align}

where $P({\bs{s}|\bs{s}^{\prime}})$ is the transition probability between states $\bs{s}^{\prime}$ and $\bs{s}$, which particular value is given by (\ref{eq:cupdate}) \cite{coolen2001}  \edit{(see Appendix \ref{a:stochastic} for details).}\par 
Given this observation, we see that an RNN can be used for the task \edit{evolving} a probability distribution $p_0$ into $p_f = L^{\targ}p_0$ by a series of \edit{stochastic} transition matrices $L^{\out}=P^{t_f}$. The goal is to adjust the parameters of the RNN (i.e. biases and connection weights) to \edit{simulate the stochastic matrix encoded in} $L^\out\approx L^\targ$ in as few steps as possible. Then, one may ask if a qRNN can do this more efficiently than any RNN. \par 
We answer this in the positive. It is worth noting that not all \edit{stochastic transition} matrices $L^{\targ}$ are embeddable in a Markov process (for a review of classical and quantum embeddability see Appendix \ref{a:stochastic}). \edit{To simulate a stochastic system’s future behavior, information about its past must be stored, and thus memory is a key resource. Quantum information processing promises a memory advantage for stochastic simulation \cite{PhysRevX.9.041013}. In simulating stochastic evolution with classical resources there is a trade-off between the temporal and physical resources needed \cite{wolpert2019space}, and it’s been shown that certain stochastic evolutions, when simulated with quantum hardware, may not suffer from such trade-off since} the evolution arising from quantum Lindbladian dynamics are far more general than classical Markovian evolution \cite{korzekwa2021quantum}. That is, there exist matrices $L^\targ$ that are quantum embeddable but not classically embeddable. Moreover, even if $L^{\targ}$ is embeddable, the quantum evolution can lower the number of steps needed to produce $L^\targ$ since the unitary dynamics of a quantum system allow a simultaneous, continuous, and coherent update of every neuron. This separation in capabilities illustrates the computational advantages of quantizing an RNN. \par  
Let us now give an example of a matrix $L^{\targ}$ that can be achieved exponentially faster in a qRNN. Consider the task of realizing a transformation $F$ corresponding to a global ``spin-flip" 
\begin{equation}\label{eq:flip}
   F_{\bs{s}|\bs{s^{\prime}}} = \begin{cases} 1 &\mbox{if } \forall_{n} \text{ } s_n\neq s^{\prime}_{n} \\
   				     0 &\mbox{otherwise.}\end{cases}
\end{equation}
Realizing $F$ on $N$ neurons using a classical Markov process requires several time steps of order $\mathcal{O}(2^{N-m})$ where $m$ is the number of hidden neurons (for details, see Sec. III.A in Ref. \cite{korzekwa2021quantum}). \edit{In other words, a classical RNN cannot produce $F$ efficiently when all available neurons must be flipped}. This is a result of (\ref{eq:cupdate}), and the fact that flipping neuron $n$ is done by ensuring that there is another neuron $m$ in the opposite state so that $J_{nm}>0$ dominates $h_n$. \par 
On the other hand, a qRNN can perform $F$ in a single step \edit{regardless if all neurons need to flip}. To see this, one can consider the case of (\ref{eq:HGeneral}) with $\Omega\gg h_{n}$. In this case, neurons both flip simultaneously and in a single time step under a unitary $U$. That is, if $|\psi_0\rangle = \sum_{\bs{s}}\sqrt{p_0(\bs{s})}|\bs{s}\rangle$, then 
\begin{equation}
    |\psi_f\rangle = U|\psi\rangle = \sum_{\bs{s}}\sqrt{p_f(\bs{s})}|\bs{s}\rangle.
\end{equation}\par
While the realization of the matrix $F$ via (\ref{eq:HGeneral}) signal a quantum advantage, we highlight that this advantage is extremely sensitive to the decoherence arising from spontaneous emission \edit{(i.e. spontaneous relaxations from $|1\rangle$ to $|\text{-}1\rangle$)}, a main source of noise in NISQ devices (see Appendix \ref{a:stochastic}). It remains an open problem whether there exist stochastic processes enabled by (\ref{eq:HGeneral}) which are robust to noise, and in the future, we hope to explore how to shield unitary stochastic processes against noise in experimentally realizable NISQ devices. \par 
The spin-flip process $F$ is efficiently simulated using a classical computer. However, $F$ exemplifies the qRNN's ability to access stochastic processes inaccessible to classical RNNs without hidden neurons. This implies that if an RNN is employed
\edit{simulate evolving $p_0$ to $p_{t_f}$ stochastically by passing it through several linear transformations, there are instances where the qRNN requires exponentially fewer steps.} \edit{Stochastic simulation, of course, has applications in finance, biology, and ecology, among other fields. As an example, Ref. \cite{blank2021quantum} used this quantum advantage to propose a quantum circuit algorithm for stochastic process characterization and presented applications in finance and correlated random walks.} The separation above illustrates the computational advantage of quantizing an RNN.

\subsection{qRNNs under spontaneous emission}\label{sec:QEOMs}
Having seen how (\ref{eq:HGeneral}) recovers the discrete update rule (\ref{eq:cupdate}), we now show that a qRNN under dissipation naturally evolves under continuous-time dynamics analogous to (\ref{eq:EOMClassical}). This establishes \edit{only} mathematical similarities between the evolution of NISQ devices and neural circuits, allowing us to use available quantum hardware for cognitive tasks, an idea that we explore further in Sec. \ref{sec:BioTasks}.\par 
Consider the qRNN in $(\ref{eq:HGeneral})$ under spontaneous emission where a spin relaxes from $|1\rangle$ to $|\text{-}1\rangle$ at a rate $\gamma$. To extract the dynamics of continuous variables, we focus on the dynamics of the expectation values of local Pauli operators.

The expectation value of an observable $A$ is $\ev{A}=\text{Tr}(A\rho)$ where $\rho$ is the density matrix describing the system. In particular, we focus on the expectations of the operators $\sigma^x_n$, and $\sigma^y_n=i|\text{-1}\rangle\langle 1|_n-i|1\rangle\langle \text{-1}|_n$. If we start the qRNN at a state for which $\langle\sigma_n^z(0)\rangle=-1$ then (see Appendix \ref{A:QEOMs})

\begin{align}\label{eq:EOMQuantum}
    \dot{\ev{\sigma^y_n}}=& -\frac{1}{\tau} \ev{\sigma^y_n(t)}-\frac{\Omega}{2\gamma}\sum_{m}J_{nm}\ev{\sigma_n^x(t)\sigma_m^y(t)}\notag\\
    &+\Delta_n(t)\ev{\sigma_n^x(t)},
\end{align}
where we have defined the neural time-scale $\tau^{-1}=\gamma/2+\Omega^2/4\gamma$ which is different than that in (\ref{eq:EOMClassical}) but bears the analogous significance of the time-scale in which $\ev{\sigma_n^y}$ decays. \par
Differently, that (\ref{eq:EOMClassical}), notice that the dynamics of $\ev{\sigma_n^y}$ are influenced by the spin's value along the $x$-axis, a consequence of the nontrivial commutation relation of spin variables. The commutation relations also make (\ref{eq:EOMQuantum}) quadratic, and therefore nonlinear. \edit{The quadratic term in (\ref{eq:EOMQuantum}) is analogous to the nonlinear term that gives RNNs their computational power.} \par 
In Appendix \ref{A:QEOMs}, we explore the dynamics of $\ev{\sigma_n^x}$ as well and show that together with $\ev{\sigma_n^y}$, we recover dynamics \edit{analogous to} integrate-and-fire RNN model \cite{Burkitt2006}, a more realistic model of neural networks in the brain than the one in $(\ref{eq:EOMClassical})$. \par 

\begin{figure}
\centering
\includegraphics[width=0.9\linewidth]{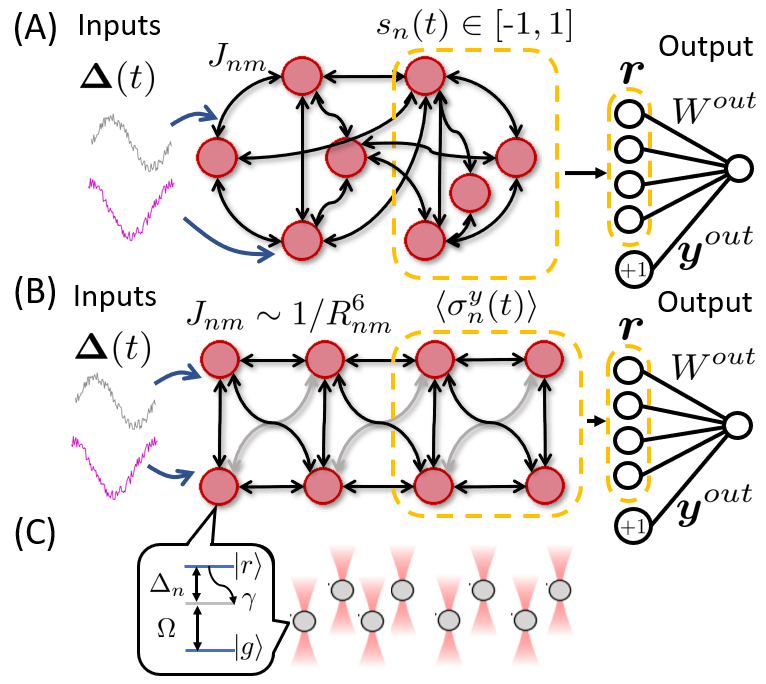}
\caption{Schematic picture of RNNs with classical and quantum neurons. \textbf{(A)} Classical RNN. The inputs are local biases, and the inter-neural connections $J_{nm}$ are arbitrary. A set of neurons is used for readout to produce the output $\bs{y}^\out = W^\out \bs{r}$. \textbf{(B)} qRNN made from Rydberg atoms which restrict the connections to $J_{nm}\sim 1/R_{nm}^6$ where $R_{nm}$ is the physical distance between atoms $n$ and $m$. \edit{Here, we depict interactions between nearest and next-nearest neighbors. However, each neuron interacts with all others in the chain via $J_{nm}\sim 1/R_{nm}^6$}. Local expectation values of a subset of atoms are for readout. \textbf{(C)} Arrays of Rydberg atoms as qRNNs. Each atom experiences a Rabi-drive $\Omega$, and a local detuning $\Delta_n$ encoding the RNN’s inputs. One of the main sources of decoherence in Rydberg atoms is spontaneous emission at a rate $\gamma$. }
\label{fig:schematic}
\end{figure}
\section{Quantum reservoir computers using Rydberg atoms: An experimental proposal}\label{sec:Rydberg}
The similarities between (\ref{eq:EOMQuantum}) and the evolution of RNNs \edit{suggest} the ability of qRNNs to \edit{emulate} neurological learning. To explore neurological learning in qRNNs, we propose to fix the architecture of the qRNN coupling constants $J_{nm}$ based on optical-tweezers arrays of Rydberg atoms. \par 
The natural Hamiltonian of a Rydberg array closely resembles the one in (\ref{eq:HGeneral}). A Rydberg atom is a single valance-electron atom that can be coherently driven between an atomic ground state $|g\rangle$ and a highly excited state $|r\rangle$ with a much larger principal quantum number. These states can represent our $|\text{-}1\rangle$ and $|1\rangle$ neuronal states respectively. A Rydberg atom in its excited state exhibits a large electronic dipole moment, and, consequently, a collection of Rydberg atoms interact via a $1/R^6$ van der Waals potential where $R$ denotes the physical distance between the two atoms. For an array of Rydberg atoms where the atoms are at fixed positions, the Hamiltonian of the system is \cite{Bernien2017}
\begin{equation}\label{eq:HRyd}
H_{Ryd} = \Delta\sum_{n} \hat{n}_n+\frac{\Omega}{2}\sum_{n}\sigma_n^x+  \sum_{nm}\frac{V}{R_{nm}^6}\hat{n}_n\hat{n}_m
\end{equation}
where $\hat{n}_n  = |1\rangle\langle 1|_n$, $\Omega$ is the coherent \edit{Rabi} drive coupling the $|\text{-}1\rangle$ and $|1\rangle$ states, $\Delta<0$ is a global drive frequency mismatch to the atomic spacing of the atoms, and $V$ is the nearest neighbor interaction strength. Using acusto-optical deflectors (AOD) and spatial light modulator (SLM), one can create spatially depending light-shifts resulting in site and time-dependent detunings $\Delta_n(t) = \Delta+\alpha(t)\Delta_n$ where $\alpha(t)$ is a time-dependent envelope. With this in mind, the Hamiltonian in (\ref{eq:HRyd}) can be mapped to a Hamiltonian like that in (\ref{eq:HGeneral}) with $J_{nm}=V/R_{nm}^6$ since $\hat{n}_n = (\sigma_n^z+\mathds{1}_n)/2$. 
\edit{In this paper, for concreteness, we compare our numerics against the experimental realization of Rydberg arrays in Ref. \cite{Bernien2017, Ebadi2020}, where the rates $\Omega, \Delta_n, V$ are all in units of mega-Hertz, while time constants are in units of micro-seconds. In these experiments, an off-resonance intermediate state, $|6P_{3/2},F=3, M_F=-3\rangle$, is used to couple $|g\rangle=|5S_{1/2}, F=2, m_F=-2\rangle$ and $|r\rangle=|70S_{1/2}, m_J=-1/2, m_I=-3/2\rangle$ of Rubidium-87 atoms through a two-photon transition.  Thus, photon-scattering off the intermediate state is the dominant source of decoherence. As we show in Appendix \ref{A:Numerics}, we can model this with a modified spontaneous emission process given by the jump operator 
\begin{equation}
    L^{+} = \sqrt{\gamma} |g\rangle\left( \alpha \langle r| + \beta \langle g|\right)
\end{equation}
instead of the typical $\sqrt{\gamma}|g\rangle\langle r|$ jump operator. In the equation above, $\gamma = 2\pi/(20 \text{ }\mu\text{s})$, and $(\alpha, \beta) = (0.05, 0.16)$ for the realistic settings we simulate.} With the full unitary and dissipative dynamics, we can think of an array of Rydberg atoms as a quantum \edit{analog} of a continuous-time RNN. Fig. \ref{fig:schematic} compares the architecture of a classical RNN in Fig. \ref{fig:schematic}A, and a Rydberg RNN in Fig.s \ref{fig:schematic}B-C. \par 
\edit{We note that training RNNs can be unstable as that often relies on (truncated) back-propagation through time or real-time recurrent learning. One way to circumvent this problem is by keeping the fixed system’s parameters. Instead, we focus on only training the output filter $W^\out$. This easier training schedule motivated the introduction of reservoir computers \cite{jaeger2002} and their quantum analogs  \cite{mujal2021opportunities,fujii2020quantum, Nakajima2019,Kutvonen2020,Ghosh2019A, khan2021physical,Gosh2019B,govia2021quantum,nokkala2021gaussian,Gosh2020,mujal2022quantum,sakurai2021quantum,PhysRevLett.127.100502}. Thus, in the following numerical experiments we fix the position of the atoms in either a 1D chain or a 2D square lattices and train only $W^\out$ and some temporal parameters depending on the task. \editt{That is, in this article we implement Rydberg reservoir computers.} Logically, successful performance on the tasks here presented sufficiently shows qRNNs computational ability. While we include the effect of small imperfections on the positions of the atoms, we see no significant effect on the performance of the tasks after averaging our results over 10 realizations of the atom's positions. We leave full optimization of the qRNN for future work.} \par 
Lastly, several features of the many-body dynamics of arrays of Rydberg atoms are particularly well suited for \edit{emulating} biological tasks. In Sec. \ref{sec:multitasking}, we show how Rydberg arrays can be used to implement inhibitory and excitatory neurons which are vital in many biological tasks such as multitasking \cite{Capano2015}. The key idea behind encoding inhibitory neurons will be leveraging positive and negative interactions between Rydberg atoms with different principal quantum numbers \cite{Han2009}. Additionally, in Sec. \ref{sec:Memory} we show that Rydberg arrays can store long-term memory by taking advantage of the weak-ergodicity breaking dynamics of quantum many-body scars \cite{Bernien2017, bluvstein2021controlling, maskara2021discrete}. \par 

\section{Learning biological tasks via reservoir computers}\label{sec:BioTasks}
We focus on analyzing the Rydberg reservoir's potential to learn biologically plausible tasks. In the tasks analyzed, we fixed the geometry of the atoms depending on the task at hand. As a proof of principle, we focus on four simple neurological tasks which indicate good performance even with a small number of atoms. We show that a Rydberg reservoir can encode inhibitory and excitatory neurons vital for successful multitasking. Likewise, we show that Rydberg reservoirs can learn to decide by distinguishing properties of stimuli, have a working memory, and exhibit long-term memory enhanced by quantum many-body scars. Simulation details of each task can be found in Appendix \ref{A:Numerics}.  \par  

\subsection{Multitasking}\label{sec:multitasking}
\begin{figure}
\centering
\includegraphics[width=\linewidth]{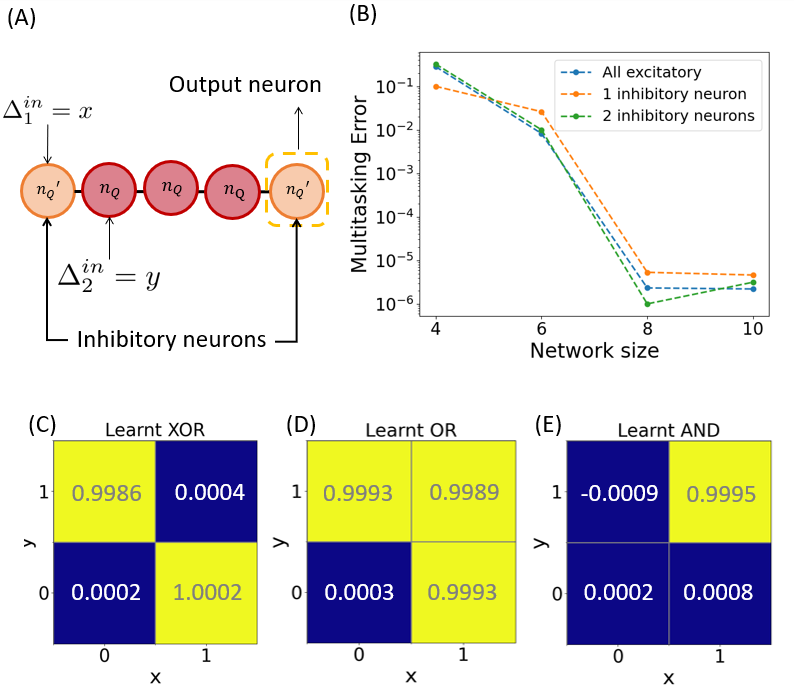}
\caption{Encoding inhibitory neurons using Rydberg atoms and using them for multitasking. Multitasking consists of fixing the qRNN's parameters and training $W^\out$ to produce three conflicting outputs. \textbf{(A)} Shows the scheme for encoding inhibitory neurons. Rydberg atoms with different principal quantum numbers are used such that pairs \edit{$(n_{Q})(n_Q')$ interact attractively while $(n_{Q}')(n_Q')$ and $(n_{Q})(n_Q)$ pairs interact repulsively}. The network receives two binary inputs $x, y$. 
\textbf{(B)} Square error for learning the functions XOR, OR, and AND on the inputs with different numbers of inhibitory neurons. Better performance is observed when 1 in every 4 neurons is inhibitory. \textbf{(C)-(E)} Example of learned functions using eight neurons and two inhibitory neurons, which results in performing 40\% better than without inhibitory neurons.}
\label{fig:multi-tasking}
\end{figure}

A hallmark of classical RNNs is their ability to multitask. Multitasking consists of simultaneously learning several output functions. Dale's principle defines an inhibitory neuron, indexed by $n$, as one with a negative sign in its interactions with all other neurons \cite{Eccles1954}
\begin{equation}\label{eq:Inhibitory}
J_{nm}\leq 0 \quad \forall m.
\end{equation}
\edit{Two Rydberg atoms with different principal quantum numbers $n_Q$, and $n_Q'$ and angular momentum quantum numbers the same can interact with a $1/r^6$ attractive potential $V_{n_Q,n_Q'}$ \cite{Han2009}. Using the Python package \texttt{PairInteraction}  \cite{Weber2017}, we note that if $n_Q$ represents the state $|r\rangle = |70S_{1/2}, m_j=-1/2, m_I=-3/2\rangle$, and $n_Q'$ represents $|r'\rangle =|73S_{1/2}, m_j=-1/2, m_I=-3/2\rangle$, then the interaction $V_{n_Q,n_Q} = V \approx -V_{n_Q,n_Q'}$ where $V$ is the strength between atoms with principal quantum numbers $n_Q$ (see Appendix \ref{A:Numerics}). We can use this fact to encode inhibitory neurons. We restrict the concentration of $n_Q'$ Rydberg atoms to be sparse such that pairs of $n_Q'$ atoms are placed as far as possible at a distance $d_{max}$ in a 1D chain arrangement. We choose the field strength $V$ so that $V/d_{max}^6= 10^{-2}$, and as a result we can neglect the interactions between pairs of $n_Q'$ atoms, but not the interactions between pairs $(n_Q)(n_Q')$ and $(n_Q)(n_Q)$. This amounts to saying that if atom $n$ is driven to $n_Q'$ then for all $m$ $J_{nm}\lesssim 0$ as in (\ref{eq:Inhibitory}). By implementing this in on our reservoir we can learn XOR, AND, and OR simultaneously for different concentrations of inhibitory neurons as illustrated in Fig. \ref{fig:multi-tasking}(A).}\par
Fig. \ref{fig:multi-tasking}(B) shows the errors of simultaneously learning XOR, OR, and AND as a function of the system size $N$ for a different number of inhibitory neurons in the array. The network is initialized in the state $|g\rangle^{\otimes N}$, and the network receives two binary inputs $x,y\in\{0,1\}$ (in units of MHz)  for a time $\Delta t$ (in units of $\mu$s) with input noise $\sigma_{in}=0.1$. Afterwards, the network is interrogated to give XOR$(x,y)$, OR$(x,y)$, and AND$(x,y)$. $W^\out$ is trained using the loss in (\ref{eq:Loss}). \edit{The errors shown in Fig. \ref{fig:multi-tasking}(B) are the minimum achieved over a wide range of choices of interaction time $\Delta t\in [0,5]$ $\mu$s.} This shows that in some cases our reservoir can benefit from having a connectivity matrix $J_{nm}$ with both positive (excitatory) and negative (inhibitory) values, analogously to the mammalian brain. \edit{For small system sizes, it seems that a ratio of 1:4 inhibitory neurons betters the learning performance, similar to the results in \cite{Capano2015}. This is supported by the performance at 4 and 8 neurons in Fig. \ref{fig:multi-tasking}(B). Particularly, $N=8$ neurons, two of which are inhibitory, result in a 40\% decrease in the loss. Nonetheless, we observe that having no inhibitory neurons is best when dealing with $N=$6 and 10 neurons. No inhibitory neurons are ever the worse choice.} Fig. \ref{fig:multi-tasking}(C)-(E) shows the results of learning XOR, OR, and AND simultaneously using $N=8$ and \edit{two} inhibitory neuron. Note that the network is fully capable of classification errors well below the input noise threshold $\sigma_{in}$.\par
\edit{Lastly, while this task shows the success of the Rydberg reservoirs at approximating Boolean functions of the input, we note that one may also want to calculate different nonlinear functions of the input. We remark that our Rydberg reservoir can approximate biologically relevant nonlinear functions such as ReLu and sigmoid.}

\subsection{Decision-making}\label{sec:decision making}
\begin{figure}
\centering
\includegraphics[width=\linewidth]{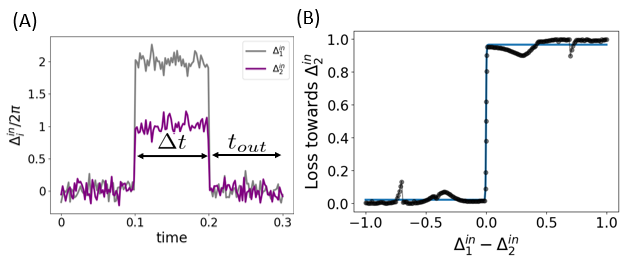}
\caption{=Decision-making task using a Rydberg reservoir. \textbf{(A)} Schematic of the input stimuli as a pair of time-dependent detunings on two atoms. The stimuli are turned on for a normally distributed time $\Delta t$ with standard deviation $\sigma_{in}=0.1$. The network decides on a relaxation time $t_{out}$ to output the decision $\text{sign}\left(\Delta^{in}_1-\Delta^{in}_2\right)$. \textbf{(B)} The psychometric response of the decision-making task which maps the accuracy towards deciding that $\Delta^{in}_{1}$ is the largest as a function of the inputs’ difference. The simulated response (dotted) is well fitted by a sigmoid function (solid curve).}
\label{fig:decision making}
\end{figure}
One of the great successes of classical RNNs is their ability to integrate sensory stimuli to choose between two actions. Here, we present the Rydberg reservoir with a variant of the dot motion decision-making task initially studied in monkeys in which several inputs are analyzed to produce a scalar nonlinear function \cite{Roitman2002}. This function represents a decision. This task shows the Rydberg reservoir’s ability to produce nonlinear functions of the input and perform simple cognitive tasks, a feature of most reservoirs proposed thus far \cite{govia2022nonlinear}. \par
In this task, a reservoir is presented with two inputs $\Delta_1^{in}$ and $\Delta_2^{in}$, and the goal is to train the network to choose which input is the largest. That is, 
\begin{equation}\label{eq:ytarg_decision}
    y^{\text{targ}} = \text{sign}\left( \Delta_1^{in}-\Delta_2^{in}\right).
\end{equation}
The stimuli, which in the case of a qRNN are local detunings on a pair of atoms, are turned on for a normally distributed time $\Delta t$ with variance also $\sigma_{in}=0.1$ and mean $\langle\Delta t\rangle=0.1$ $\mu s$ (see Fig. \ref{fig:decision making}(A)). The stimuli are then turned off, and the network chooses a relaxation time $t_{out}$ after which it ``makes a decision" by approximating (\ref{eq:ytarg_decision}). This is known as the fixed-duration protocol since the experimentalist fixes the stimulation period, and the subject, the reservoir in this case, learns to choose a response time $t_{out}$. \par
In the brain, we expect the performance of a decision-making task to follow a sigmoidal psychometric response \cite{Song2016_ExInh,Roitman2002}. A psychometric response maps out the accuracy of a decision-making task as a function of stimuli distinguishability. As an example of a psychometric response, the reader could think about paying a routine visit to the eye doctor and having to discern the letters ``b" and ``p" written on the wall. If the letters are large enough, they become distinguishable, and if the letters are too small one often fails to make out the right letter. \par
\edit{Classically, a decision-making task benefits from connectivity between all neurons. Since our connectivity is limited by physical constraints, a 2D square lattice structure was chosen to prevent neurons from being isolated from the rest. Moreover, a 2D square lattice is experimentally friendly.} We set up a Rydberg reservoir of $3 \times 2$ atoms with two input atoms and two different output atoms (for details see Appendix \ref{A:Numerics}). The reservoir is then trained by optimizing over $t_{out}$, and $W^\out$ such that the reservoir’s output approximates (\ref{eq:ytarg_decision}) \edit{while keeping the network parameters $J_{nm}$, $\Omega$ and $\Delta^{in}_n$ fixed.} We observe that $t_{out}\approx 1$ $\mu$s is regularly obtained as this is the time scale in which the information about $\Delta_{1,2}^{in}$ propagates through the network. In our case, $c_1 = \Delta^{in}_1-\Delta^{in}_2$ is a natural choice for a measure of stimuli distinguishability. Fig. \ref{fig:decision making}(B) shows the psychometric response of the task which is qualitatively similar to the ones obtained in classical RNNs \cite{Song2016_ExInh}. Moreover, we see in Fig. \ref{fig:decision making}(B) that if \edit{$|c_1|\geq \sigma_{in}$ such that it is above the input noise level} our network success more than $80\%$ of the time. The success of this task shows the Rydberg reservoir’s ability to \edit{emulate} simple cognitive tasks. \par

\subsection{Parametric working memory}
\begin{figure}
\centering
\includegraphics[width=\linewidth]{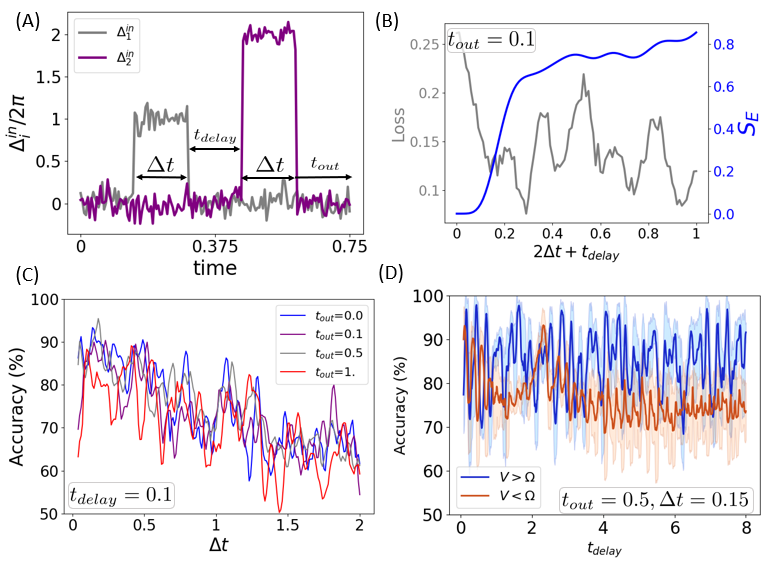}
\caption{\edit{Working memory of a Rydberg quantum reservoir computer. \textbf{(A)} Schematic of the network's inputs where two atoms are detuned for a time $\Delta t$ but temporally separated by a time $t_{delay}$. Two different output neurons are used for readout at a time $t_{out}$ after the second input is turned off. \textbf{(B)} Loss of the working memory task as a function of the total input time $2\Delta t+t_{delay}$ (gray). Entanglement entropy between the input qubits and the rest of the reservoir as a function of $2\Delta t+t_{delay}$ (blue). Here, the mean value of $t_{out}$ is 0.1. The loss stays large for small input times until the input qubits start entangling with the rest of the reservoir. (C) Accuracy as a function of the time the input are turned on ($\Delta t$) for four different choices of $t_{out}$ and with fixed $t_{delay}=0.1$. These curves show that accuracy is largely independent of $t_{out}$ and $\Delta t$ as long as $\Delta t<0.3$ (D) Accuracy of the working memory task at $\Delta t=0.15$ and $t_{out}=0.5$ as a function of $t_{delay}$. The blue curve is the performance when $V>\Omega$ puts the reservoir in the Rydberg blockaded regime, while the red curve is the performance when $V<\Omega$ puts the reservoir in the disordered regime. These plots show that when $V> \Omega$, the Rydberg reservoir can hold memory for later manipulation better than when $V< \Omega$. Shaded regions indicate error bars.}
}
\label{fig:working-memory}
\end{figure}
Our next neurological task is that of parametric working memory. Working memory, which is one of the most important cognitive functions, deals with the brain's ability to retain and manipulate information for the later execution of a task. Here, we train a network to perform a task based on the decision-making task in Sec. \ref{sec:decision making} but with two temporally separate stimuli (see Fig.  \ref{fig:working-memory}(A)). We use the fixed-time protocol where the separation between stimuli, denoted by $t_{delay}$ if fixed by us. The stimuli are both turned on for a time $\Delta t$, \edit{and after the second input the network is left to relax for a time $t_{out}$ before two output neurons are used to approximate (\ref{eq:ytarg_decision}. To avoid overfitting, we add Gaussian noise to the times $\Delta t$, $t_{out}$, and $t_{delay}$ with zero mean and standard deviation $\sigma_{in}=0.1$. The network optimizes over $W^\out$.} Thus, the network has to retain information about $\Delta_1^{in}$ for a few ``seconds" to then compare against $\Delta_2^{in}$ and make a decision. \par 
We set a Rydberg reservoir of $3 \times 2$ atoms with two input atoms and two different output atoms (for details see Appendix \ref{A:Numerics}). \edit{Fig. \ref{fig:working-memory}(B) shows the loss of the network as a function of the total time the inputs are injected into the network ($\tau=2\Delta t+t_{delay}$). We note that the loss function is high for small $\tau$ since it takes the input neurons to correlate with the rest of the reservoir. Accordingly, in Fig. \ref{fig:working-memory}(B) we show that growth of the entanglement entropy of the input qubits accompanies a decrease in the loss function. For Fig.   \ref{fig:working-memory}(C) we fixed $t_{out}=0.1$, a choice which has little effect on the reservoir's performance. \par 
In Fig. \ref{fig:working-memory}(C) we show the accuracy of the reservoir at reproducing (\ref{eq:ytarg_decision}) as a function of the time the inputs are turned on ($\Delta t$) and for different choices of $t_{out}$. For these plots $t_{delay}=0.1$ is fixed. We notice that the accuracy is largely invariant to our sampled choices of $t_{out}$. \par 
Lastly, in Fig. \ref{fig:working-memory}(D), we probe the reservoir's accuracy as a function of $t_{delay}$. For these experiments, we fix $t_{out}=0.5$ and $\Delta t=0.15$. Importantly we set $V=2\pi \times 10$ MHz and $\Omega= 2\pi\times 4.2$ MHz such that $V> \Omega$ and neighboring Rydberg excitations are off-resonance putting our reservoir in the so-called \textit{blockaded regime} \cite{urban2009observation, gaetan2009observation}. While one initially might expected the accuracy to decrease for increasing $t_{delay}$, we found that this is not the case and instead the accuracy oscillates persistently reaching high accuracies as shown in Fig. \ref{fig:working-memory}(D) blue curve. Interestingly, this behavior disappears when the coupling $V=2\pi \times 0.1$ MHz such that $V< \Omega$ as shown in the red curve in Fig. \ref{fig:working-memory}(D), although the performance is statistically significant even for long $t_delay$ with an accuracy greater than $50\%$. We can conclude that, in the blockaded regime, the reservoir can hold information for longer periods. We can understand this dependence on $V/\Omega$ as follows. In the disordered regime, atoms are mostly uncorrelated and the atoms are allowed to freely oscillate with the dynamics being dominated by the drive $\Omega$. Thus, after a short time, the inputs coming through a $z$-field are largely irrelevant and the network is unable to hold the information about the first input. On the other hand, when $V>\Omega$ the atoms are largely correlated since neighboring excitations of Rydberg atoms are blockaded and the dynamics are slowed down. These slow dynamics in the system allow for longer memory times. In Sec. \ref{sec:Memory}, we will explore the longer-term memory in the blockaded regime and show that long-term memory in a reservoir can be stabilized due to the presence of quantum many-body scars.} 
\subsection{Long-term Memory via Quantum Many-body Scars}\label{sec:Memory}
\begin{figure}
\centering
\includegraphics[width=0.90\linewidth]{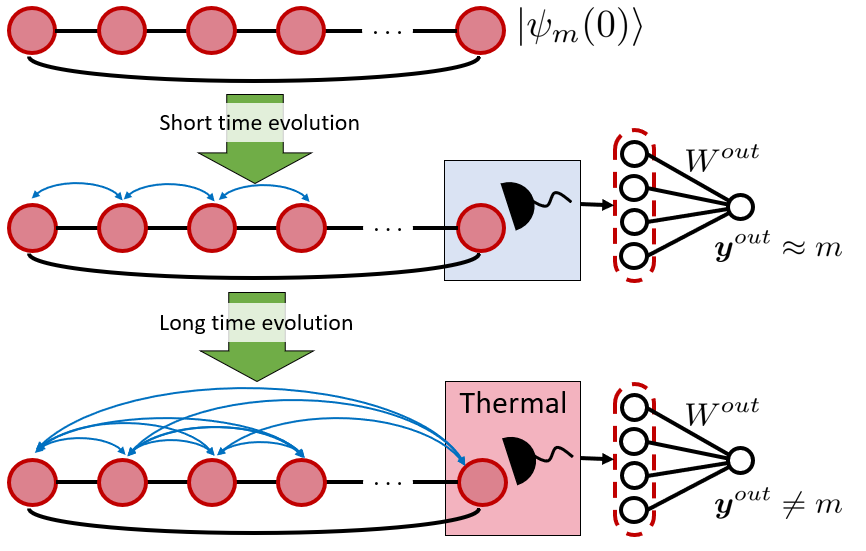}
\caption{A state encoding a memory $m$ is prepared. The state evolves under its natural Hamiltonian before being interrogated via local measurements to retrieve $m$. If the evolution time is short, the system is yet out of equilibrium and remembers its initial condition. Thus, $m$ can be retrieved. On the other hand, after a long time, the system may thermalize and local measurements fail to provide information about the initial state. Thus, the memory retrieval time is upper bounded by the thermalization time of the initial state $|\psi_m(0)\rangle$ under the system's dynamics. In the example in Sec. \ref{sec:Memory}, the system is a chain of Rydberg atoms, and final measurements are performed on a single atom which is then linearly post-processed to retrieve $m$. In this case, a thermal state can be observed by measuring if the entanglement entropy of the region obeys a volume law. If the dynamics can be stabilized against thermalization, the memory can be retrieved at larger times.}
\label{fig:thermalization}
\end{figure}
Finally, we turn to examine a reservoirs ability to encode long-term memory. The task consists of encoding a classical bit $m$ in the initial state of a reservoir $|\psi_m(0)\rangle$ so that after the system is left to evolve under its inherent dynamics for a time $T$, local measurements of the state $|\psi_m(T)\rangle$ are used to recover $m$. However, $m$ cannot be recovered from local measurements if the dynamics obey the eigenstate-thermalization hypothesis (ETH) \cite{d2016quantum}. Instead, local measurements of $|\psi_m(T)\rangle$ obey thermal statistics described by the energy spectrum of the Hamiltonian and bear no information on the initial condition $|\psi_m(0)\rangle$. Thus, reservoirs that violate the ETH are naturally suited for memory tasks, since they can locally retain information about their initial state. Indeed this notion has begun to be studied in quantum reservoirs \cite{sakurai2021quantum, PhysRevLett.127.100502}. Recent experiments using quench dynamics in arrays of Rydberg atoms have revealed quantum many-body scaring behavior \cite{Bernien2017}, which can be stabilized \cite{bluvstein2021controlling, maskara2021discrete} to delay the thermalization of the system. Here, we use these results to enlarge the memory lifetime of a reservoir. Simulation details are found in Appendix \ref{A:Numerics}. \par
In the case of a kicked ring of Rydberg atoms experiencing nearest-neighbor blockade, the dynamics are captured in the so-called PXP-model \cite{Bernien2017, maskara2021discrete, fendley2004competing, lesanovsky2012interacting}
\begin{align}\label{eq:PXP1}
    H(t) &= H_{PXP} + \hat{N}\sum_{k\in \mathbb{Z}}\theta_k\delta(t-k\tau) \\
    H_{PXP} &= \Omega\sum_{n=1}^{N}P_{n-1}\sigma_n^xP_{n+1} \qquad \hat{N} = \sum_{n}\hat{n}_n
\end{align}
where $P_{n} = |g\rangle\langle g|_n$ projects the atom at the $n^{th}$ site onto the ground state \edit{and we choose periodic boundary conditions to mitigate edge effects}. In (\ref{eq:PXP1}) we let $\theta_k = \pi+\epsilon_k$ where $\epsilon_k$ is a Gaussian random variable with mean $\epsilon$ and variance $\sigma_{in}^2$. That is $\epsilon_k$ plays the role of added noise in the reservoir. \edit{For this discussion we let $\gamma=0$ since we know from experiments that the quantum scaring behavior is robust to the atom's decoherence, and the choice to work with the Hamiltonian evolution helps speed up the acquisition of numerical data.}\par 
We denote $\chi_\tau = \exp{(-i\pi \hat{N})}\exp{(-i\tau H_{PXP})}$. It has been empirically observed that $\chi_\tau$ approximately exchanges the Neel states $|AF\rangle = |1010...\rangle$ and $|AF'\rangle = |0101...\rangle$ for $\tau \approx 1.51\pi \Omega^{-1}$ \cite{Bernien2017}. Note that $\chi_\tau \chi_\tau = \mathds{1}$, and so under no noise, any state $|\psi\rangle$ is recovered after a cycle of evolution of $2\tau$. However, the noise $\epsilon_{k}$ destroys the revival of all initial states except for  $|AF\rangle$ and $|AF'\rangle$ (see Appendix \ref{A:QScars}). This leads to many-body quantum scars stabilized by the operator $\exp{(-i\pi \hat{N})}$ \cite{bluvstein2021controlling, maskara2021discrete}. \par

Given the dynamics in (\ref{eq:PXP1}), we propose the following scheme for encoding a binary memory $m\in \{0,1\}$. We choose a reference state $|\psi\rangle$, and let $|\psi_0(0)\rangle= |\psi\rangle$ and $|\psi_1(0)\rangle = \chi_\tau |\psi\rangle$. Subsequently, the state $|\psi_m(0)\rangle$ is left to evolve for $n$ cycles of duration $2\tau=2(1.51\pi)$ after which the populations $\bs{r}_m(n) = (P_g(2n\tau|m), P_r(2n\tau|m))$ of the single-atom reduced density matrix are used to retrieve $m$. The retrieval is done by training a vector $W^\out_{n}$ on $M$ instances of $\bs{r}_m(n)$ in order to minimize (\ref{eq:Loss}) with $\bs{y}^\text{targ} = \bs{m}$ the binary vector of memories and $\bs{y}^\out(n) = W_n^\out \bs{r}(n)$ our networks' output after $n$ cycles. \par
\begin{figure}
\centering
\includegraphics[width=0.90\linewidth]{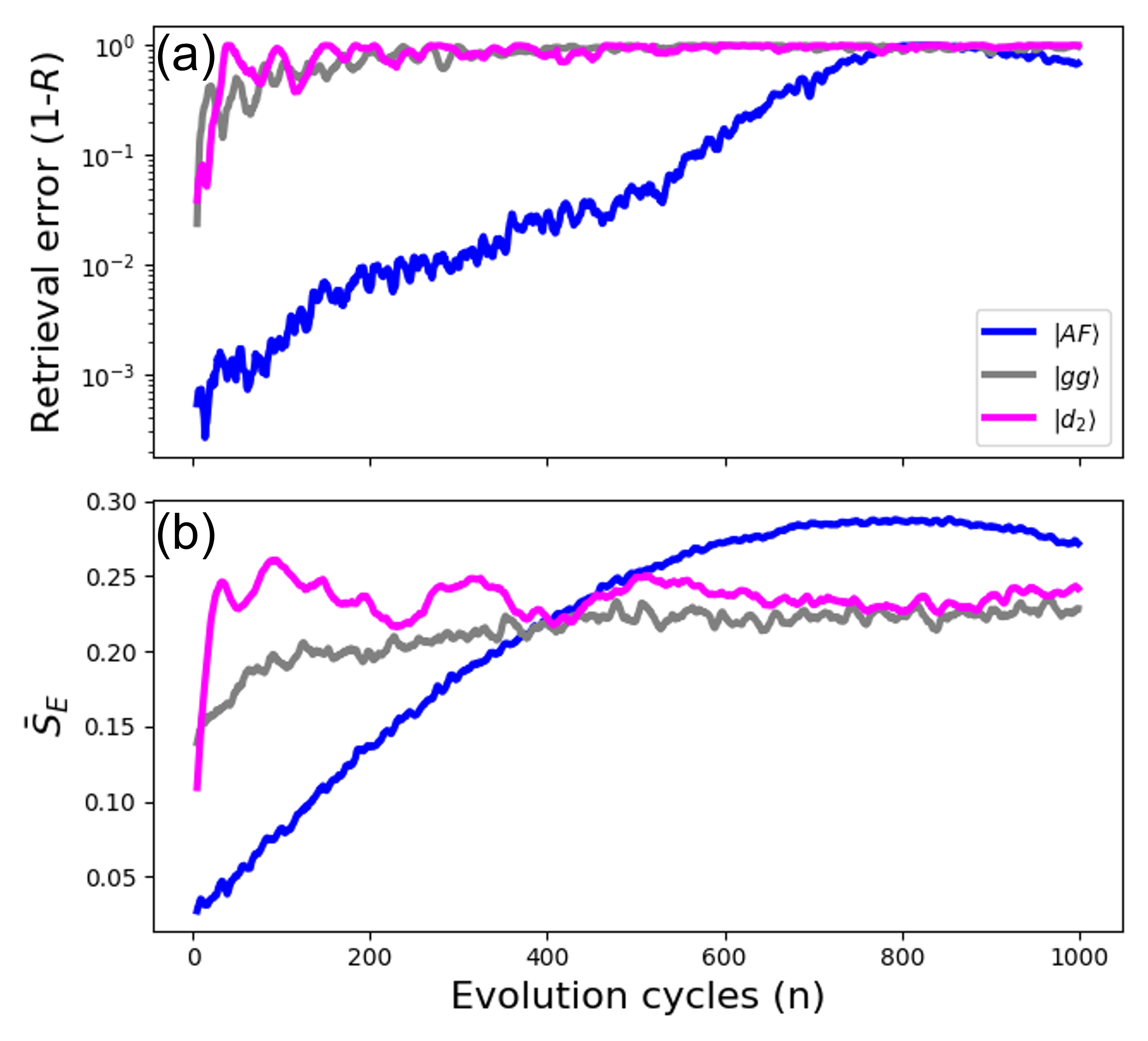}
\caption{Dependence of memory retrieval on different reference states. We use a ring of $N=8$ Rydberg atoms with $\epsilon=\sigma=0.1$, and $M=100,30$ samples for the training and testing sets respectively. The memories are sampled from a balanced Bernoulli distribution. \textbf{(a)} Shows the memory retrieval error for three different choices of reference state, $|AF\rangle = |grgrgrgr\rangle$, $|gg\rangle = |gg...g\rangle$ and $|d_2\rangle = |grggggrg\rangle$. Due to the scaring behavior of $|AF\rangle$, the memory length is greatly improved. \textbf{(b)} Shows the left-most atom's entanglement entropy averaged over the $M$ memory instances ($\bar{S}_E$). Saturation of $\bar{S}_E$ signals the thermalization of the system and thus a decrease in $R$.}
\label{fig:Memory}
\end{figure}

To quantify the quality of the memory retrieval $R(n)$, we use the squared Pearson's $r$-factor
\begin{equation}
    R(n) = \frac{\text{cov}^2(\bs{m}, \bs{y}^{\out}(n))}{\sigma^2(\bs{m})\sigma^2(\bs{y}^\out(n))}.
\end{equation}
Fig. \ref{fig:Memory}(a) shows the memory retrieval error as a function of the number of cycles for three different choices of reference states. Fig. \ref{fig:Memory}(b) shows the average entanglement entropy $(\bar{S}_E)$ of the left-most atom in the ring. Saturation of $\bar{S}_E$ signals growth in the memory retrieval error as the state ``forgets'' the initial condition. From other studies, we see that memory is retrieved at longer times due to the slow thermalization of the Neel states due to quantum many-body scars \cite{Bernien2017, maskara2021discrete, fendley2004competing, lesanovsky2012interacting,bluvstein2021controlling, maskara2021discrete}. The time-crystalline nature of the reservoir using $|\psi\rangle = |AF\rangle$ signals long-time correlations, and thus the reservoir can be used to encode and predict series with long-time correlations \cite{Kutvonen2020}.\par 
The Neel states exhibit long-term memory due to the evolution's scaring behavior. This can be understood by analyzing the average evolution produced by a single cycle. Up to second order in $\epsilon_k$, the state at time $2\tau n$, $\rho(n)$, evolves to the state at time $2\tau(n+1)$, $\rho(n+1)$, where (see Appendix \ref{A:QScars})
\begin{align}\label{eq:Lind}
    \rho(n+1) &= \rho(n)-i\epsilon[H^{+}, \rho(n)]\notag\\
    &+\sigma^2_{in}\left(H^{+}\rho(n)H^{+} -\frac{1}{2}\{H^{+}H^{+}, \rho(n)\}\right)\notag \\
    &+\sigma^2_{in}\left(H^{-}\rho(n)H^{-} -\frac{1}{2}\{H^{-}H^{-}, \rho(n)\}\right).
\end{align}
Here, $H^{\pm} = \hat{N}\pm \chi_\tau \hat{N}\chi_\tau$ are Hermitian operators. We can rewrite (\ref{eq:Lind}) as $\rho(n+1) = \rho(n)+\mathcal{L}_{\epsilon,\sigma}(\rho(n))$. Since $[H^{+}, \chi_\tau] =0$, the operator $H^{+}$ has an emergent $\mathbb{Z}_2$ symmetry which means that the ground states of $H^{+}$ are well approximated by the states $|\pm\rangle = \frac{1}{\sqrt{2}}\left(|AF\rangle\pm|AF'\rangle\right)$ \cite{maskara2021discrete}. Note that 
\begin{align}
    H^+|+\rangle &\approx N|+\rangle, \quad &&H^-|+\rangle \approx 0,\\
    H^+|-\rangle &\approx N|+\rangle, \quad &&H^-|-\rangle \approx 0,
\end{align}
were $N$ is the system size We conclude that if $\rho(n)=|AF\rangle\langle AF|$  then $\rho(n+1)\approx\rho(n)$ as this state is (approximately) in the kernel of $\mathcal{L}_{\epsilon,\sigma}$. Therefore, the Neel states are suitable memory states. \par
Equation (\ref{eq:Lind}) also tells us that any density matrix in the kernel of $\mathcal{L}_{\epsilon,\sigma}$ may also serve as a memory state since it is a steady state of the evolution. This would allow us to enlarge the number of memories accessible in a qRNN. In Appendix \ref{A:QScars} we show the existence of a large number of steady states, and we present a scheme to prepare a number of them. It's worth noting, however, that these memories may have to be distinguished from one another via global measurements. The questions of how to efficiently prepare and distinguish these memory states \edit{remain importantly both open and key in telling us if a memory quantum advantage can be claimed in qRNNs. As it stands, using quantum scars signals that Rydberg-inspired RNNs may present enhanced memory since quantum scars are classically simulatable due to their low entanglement entropy. However, it's unclear whether the system can be classically simulated at late times due to the onset of the thermalization. These questions are left for future studies.}\par 
Quite recently, another proposal to enlarge the number of memories accessible in a quantum reservoir has been introduced using the emergent scale-free network dynamics of a melting discrete time-crystal in an Ising chain \cite{sakurai2021quantum}. The proposal in \cite{sakurai2021quantum} can be seen as a generalization of the quantum reservoir presented in (\ref{eq:PXP1}) by dropping the constraint of the Rydberg blockade. Our results, as well as those in \cite{sakurai2021quantum}, pose the possibility of having an RNN with a memory capacity that outpaces that of classical RNNs such as the Hopfield network \cite{folli2017maximum}.

\section{Conclusions and outlook}\label{sec:Conclusion}
In this article, we present a quantum extension of a classical RNN on binary neurons. This implies a deep connection between controllable many-body quantum systems and brain-inspired computational models. Our qRNN facilitates the ability to employ the analogue dynamics of quantum systems for computation instead of the circuit-based paradigm. We show how features of the quantum evolution of our qRNN can be used for quantum learning tasks, and to speedup \edit{simulation of stochastic dynamics}. We implement a quantum reservoir using arrays of Rydberg atoms and show how Rydberg atoms \edit{analogously} perform biological tasks even in the presence of a few atoms. This can be explained via the physics of the system. For example, we showed how weak-ergodicity breaking collective dynamics in Rydberg atoms can be employed for long-term memory. \par
While this article takes the first step forward in connecting controllable quantum systems and neural networks from a fundamental perspective, several questions remain unanswered. 
\edit{Firstly, from the first two quantum features hereby presented, studies of how qRNNs can be used for quantum error correction in circuit-like quantum computing are warranted.}
Directly from this work, investigations into advantageous stochastic processes in qRNNs that are robust to decoherence are enticing. These advantages will likely emerge from the collective behavior of quantum neurons. Therefore, the field will soon require a thorough understanding of the collective dissipative dynamics of neurons in qRNNs, which would also shed light on rigorous studies of the computational power of these architectures. Guided by the fact that neural networks become universal approximators by interconnecting many neurons, one may also consider the spatial and control requirements necessary for universal brain-inspired quantum machine learning.\par
Given the vast number of classical computational models for the brain, there are several immediate research directions. One of these is the exploration of a systematic way to quantize more biologically realistic models of a neural circuit. A possible starting point for translating different neural circuits would be to exploit key engineering and fundamental features of different NISQ platforms. For example, recent experiments using Rydberg atoms in photonic cavities may provide us with the ability to capture neural plasticity on qRNNs by arbitrarily tuning the inter-neural interactions \cite{periwal2021programmable}.  Likewise, superconducting circuits have lately been used to encode biologically realistic single-neuron models \cite{gonzalez2020quantized}. Along these explorations, it will be imperative to establish a variety of methods to analyze how quantum neural networks recover the classical protocols within certain limits, as well as the source and extent of the quantum advantages that each platform can offer. \par 
Lastly, while our memory encoding scheme in Sec. \ref{sec:Memory} offers a possibility to encode a binary memory, whether a higher number of memories can be encoded efficiently remains an important open question. In Appendix \ref{A:QScars} we offer a proposal based on the steady states of the effective dissipative evolution in the pre-thermalization regime introduced by the noise in the qRNN. This already shows a theoretical number of memories greater than those attainable by the vanilla Hopfield network \cite{folli2017maximum}. However, distinguishing these memories, or producing Hamiltonians with the desired memory state in mind, is left for future research. It is clear, however, that memory in a quantum reservoir relies on ergodicity breaking dynamics \cite{sakurai2021quantum,PhysRevLett.127.100502}. Hamiltonian engineering techniques, together with more general driven Hamiltonians such as those in \cite{sakurai2021quantum}, may pave the way towards programmable memories in a qRNN.

\section*{ACKNOWLEDGMENTS}
The authors thank Mikhail. D. Lukin and Nishad Maskara for insightful discussion. RAB acknowledges
support from NSF Graduate Research Fellowship under Grant No. DGE1745303, as well as funding from Harvard University's Graduate Prize Fellowship. 
XG acknowledges
support from Quantum Science of the Harvard-MPQ Center for Quantum Optics, the Templeton Religion Trust grant TRT 0159, and by the Army Research Office under Grant W911NF1910302 and MURI Grant W911NF-20-1-0082.
SFY acknowledges funding from NSF and AFOSR.

\appendix
\renewcommand{\theequation}{A.\arabic{equation}}
\bibliography{actual}

\section{Probability transformations using qRNNs}\label{a:stochastic} 
\edit{In the case of of the RNN presented in (\ref{eq:cupdate}), using Ref. \cite{coolen2001} we can derive that $P(\bs{s}|\bs{s}')$ in (\ref{eq:Markov}) is given by
\begin{equation*}
    P(\bs{s}|\bs{s}') = \prod_{i=1}^N \frac{1}{2}\left(1+s_ig\left[h_i(\bs{s}')/\sigma^2_{in}-1\right]\right)
\end{equation*}
where $g\left[x\right] = \text{Erf}\left[z/\sqrt{2}\right]$ is the error function due to the Gaussian noise. Regarding the task in Sec. \ref{sec:QAstochastic} of flipping all neurons at once, one could naively think that this can be done classically by taking the inputs $\Delta_n\rightarrow \infty$, however, since the noise's strength $\sigma_{in}^2$ scales as the size of the inputs, one obtains $P(\bs{s}|\bs{s}') \rightarrow \prod_{i=1}^N\frac{1}{2}(1+s_i/2)$ which is a completely random update independent of the original state.}\par
A transition matrix $L$ obeys $L_{\bs{s}'|\bs{s}}\geq0$ and $\sum_{\bs{s}'}L_{\bs{s}'|\bs{s}}=1$. $L$ is said to be \textit{classically embeddable} if it can be generated by a continuous Markov process via
\begin{equation}
\frac{d}{dt}P(t) = K(t)P(t), \quad P(0)=\mathds{1}, \text{ } P(t_f)=L,
\end{equation} 
where $K$ is called a generator matrix that preserves the positive nature of $P$ via the constraint $K_{\bs{s}|\bs{s}^{\prime}}\geq 0$ for $\bs{s}\neq \bs{s}^{\prime}$, and normalization via the constraint $\sum_{\bs{s}}K_{\bs{s}|\bs{s}^{\prime}}=0$. Applied to our setup, a classically embeddable stochastic process is one that can transform $p_{t_f}=L p_{0}$ via an RNN without employing any hidden neurons (i.e. $M=N$ neurons are used for readout), and in a single step. In general, determining if a matrix $L$ is embeddable is an open question, but any embeddable matrix must necessarily satisfy \cite{goodman1970intrinsic}
\begin{equation}\label{eq:cembe}
    \prod_{\bs{s}}L_{\bs{s}|\bs{s}}\geq \text{det}L\geq 0.
\end{equation}
From (\ref{eq:cembe}), it immediately follows that the global ``spin-flip" matrix $F$ defined in (\ref{eq:flip})
is not classically embeddable. That is, $\det F=1$ and $\prod_{\bs{s}}F_{\bs{s}|\bs{s}}=0$, violating (\ref{eq:cembe}). Notice that the impossibility of performing $F$ without hidden neurons is quite general, and it is not limited to the stochastic process allowed by (\ref{eq:cupdate}). Moreover, the number of time-steps needed to achieve $F$ using $m$ hidden neurons is of order $\mathcal{O}(2^{N-m})$ (for details, see Sec. III.A in Ref. \cite{korzekwa2021quantum}). \par
Similar definitions of embeddability exist in the quantum setting. A stochastic process $L$ is said to be \textit{quantum embeddable} if there exists a Markovian quantum channel $\mathcal{E}$ such that 
\begin{equation}
    L_{\bs{s}'|\bs{s}} = \langle \bs{s}'|\mathcal{E}(|\bs{s}\rangle\langle\bs{s}|)|\bs{s}'\rangle.
\end{equation}
A Markovian quantum channel $\mathcal{E}$ is a channel arising from the time-evolution under a master equation, and thus $\mathcal{E}$ may include unitary and dissipative terms.  As pointed out in Ref. \cite{korzekwa2021quantum}, all classically embeddable stochastic processes. Moreover, permutations such as $F$ in (\ref{eq:flip}) are quantum embeddable since all permutations are unitary operators.\par 
We highlight that realizing $F$ is extremely sensitive to the decoherence arising from spontaneous emission, a main source of noise in NISQ devices. If $\gamma$ is the decay rate at which spin $|1\rangle$ relaxes to $|\text{-}1\rangle$, one can show that the unitary evolution leads to the stochastic process $F^\gamma$ where $\text{det}F^\gamma=e^{-\mathcal{O}(2^N)}$. Notice that whether $F^\gamma$ violates (\ref{eq:cembe}) becomes rapidly inconclusive with increasing system size.

\section{Continous-time dynamics for a qRNN}\label{A:QEOMs}
\renewcommand{\theequation}{B.\arabic{equation}}
A successful neural circuit model is the integrate and fire RNN (IF-RNN). In an IF-RNN each of the $N$ neurons is influenced by pre-synaptic firing rates and produces a post-synaptic firing rate as an output. Each neuron is endowed with a firing rate $s_{n}(t)$, where $n$ denotes the $n^{th}$ neuron. The pre-synaptic firing rates arriving at the $n^{th}$ neuron are integrated to produce a pre-synaptic current  $I_n(t)$.  In turn, the neuron produces a firing rate $s_n$ influenced by its current and the firings of other neurons. Additionally, each neuron can receive a temporal input stimulus $\Delta^{in}_n(t)$ which affects both the currents and the firing rates. Generally, the firing rates and currents are described by non-linear, coupled differential equations of the form
\begin{align}
    \dot{I}_n &= -\tau_I^{-1}I_{n}+G_n(\bs{s}(t), \bs{I}(t), J_{nm}, \bs{\Delta}^{in}(t)) \label{eq:I}\\
    \dot{s}_n &= -\tau^{-1}_s s_{n}+F_n(\bs{s}(t), \bs{I}(t), J_{nm}, \bs{\Delta}^{in}(t)) \label{eq:r}
\end{align}
where $\tau_{I,r}$ are relaxation time constants for the currents and firing rates respectively. The vector $\bs{s}(t)$ is defined as $\bs{s}(t) = (s_1(t),...,s_N(t))$, with $\bs{I}(t)$, and $\bs{\Delta}^{in}(t)$ defined analogously. The functions $G$ and $F$ ensure the dynamics are non-linear which gives RNNs their vast computational complexity. The specific forms of $G$ and $F$ depend on the application and relation between the currents and the firing rates one is trying to capture by the model.\par
The qRNN in Sec. \ref{sec:QEOMs} follows
the Heisenberg-Langevine equations of motion 
\begin{align}\label{eq:Louvillian}
    \dot{A} = i[H,A]&+\sum_{n}\left(\frac{\gamma}{2}\sigma_n^{+} + f_n^\dagger\right)[A,\sigma_n^{-}]\notag\\
    &+\sum_{n}[A,\sigma_n^{+}]\left(\frac{\gamma}{2}\sigma_n^{-} + f_n\right)
\end{align}
for any operator $A$. In (\ref{eq:Louvillian}), $\sigma_n^{+} = |1\rangle\langle\text{-}1|_n$, $\sigma^{+}_n = (\sigma_n^{+})^\dagger$, and $f_n$ is a Langevin noise operator with Gaussian statistics $\ev{f_n(t)}=0$ and $\ev{f_n(t)f_m^\dagger(t')} \propto \delta_{mn}\delta(t-t')$. In the equation above $[A,B] \equiv AB-BC$ stands for the commutator between matrices $A$ and $B$.\par
To extract the statistics of the system, one may choose to look at the dynamics of two different local observables' expectation values. For example, the equations of motion for expectations of the local Pauli operators $\sigma^x_n =|\text{-1}\rangle\langle 1|_n+|1\rangle\langle \text{-1}|_n$, and $\sigma^y_n=i|\text{-1}\rangle\langle 1|_n-i|1\rangle\langle \text{-1}|_n$ are given by 
\begin{align}
    \dot{\ev{\sigma^x_n}} = -\frac{\gamma}{2}\ev{\sigma^x_n} +i\ev{\left[H(t),\sigma_n^x\right]} \label{eq:sigmax}\\
    \dot{\ev{\sigma^y_n}} = -\frac{\gamma}{2}\ev{\sigma^y_n} +i\ev{\left[H(t),\sigma_n^y\right]} \label{eq:sigmay}
\end{align}
with $H(t)$ specified by (\ref{eq:HGeneral}). The expectation values are calculated in the quantum-mechanical sense such that for an operator $A$, $\ev{A}=\text{Tr}(A\rho)$, and terms linear in $f_n$ cancel out. Notice that the commutators in equations (\ref{eq:sigmax})-(\ref{eq:sigmay}) play the role of the functions $G$ and $F$ in (\ref{eq:I})-(\ref{eq:r}).\par
For $\sigma_n^z$, (\ref{eq:Louvillian}) gives 
\begin{equation}\label{eq:sigmaz}
    \dot{\sigma^z_n} = -\gamma/2\sigma^z_n-\frac{\Omega}{2}\sigma_n^y+\gamma \mathds{1}/2-2f_n^\dagger \sigma_n^{-}.
\end{equation}
This can be integrated out to give
\begin{align}
    \sigma_n^z(t)-\sigma_n^{z}(0) = &\int_{0}^t dt'e^{-\gamma/2(t-t')}\biggl(-\frac{\Omega}{2}\sigma_n^y(t') \notag \\
    &-2f_n^{\dagger}(t')\sigma_n^{-}(t')+\gamma/2\mathds{1}\biggr)
\end{align}
We choose to start the network at $\ev{\sigma^z_n}=-1$ for all $n$. We plug this back into (\ref{eq:sigmay}), and we take the expectation values to eliminate terms linear in $f_n$. We obtain
\begin{align}\label{eq:RydMemoryEOMs}
    \dot{\ev{\sigma^y_n}} =& -\frac{\gamma}{2}\ev{\sigma_n^y}+\frac{\Omega}{2}\sum_{m=1}^N J_{nm}\int_{0}^t dt' e^{-\gamma(t-t')}\ev{\sigma^x_n(t)\sigma^y_m(t')} \notag\\
    & +\Delta_n(t) \ev{\sigma_n^x} - \frac{\Omega^2}{4}\int_{0}^{t}\ev{\sigma^y_n(t')}e^{-\gamma(t-t')}dt'.
\end{align}
Similar equations can be found for $\ev{\sigma_n^x}$. Equation (\ref{eq:RydMemoryEOMs}) tells us that $\ev{\sigma_n^y}$ depends on past statistics, and thus our network has a memory time bounded by $1/\gamma$. Let $J$ denote the matrix $J_{nm}$. For \edit{$\gamma t\gg 1$}, we can extend the lower bound of integration to $-\infty$. Using the approximation $\int_{-\infty}^t e^{-\gamma(t-t')}f(t')dt'\approx -\gamma^{-1}f(t)$, we obtain
\begin{align}
    \dot{\ev{\sigma_n^x}} &= -\frac{\gamma}{2}\ev{\sigma_n^x} - \Delta_n^{in}\ev{\sigma_n^y} \notag\\
    &-\frac{\Omega}{2\gamma}\sum_{m}J_{nm}\ev{\sigma_n^y \sigma_m^y} \label{eq:sigmaxSimp} \\
    \dot{\ev{\sigma_n^y}} &= -\left(\frac{\gamma}{2}+\frac{\Omega^2}{4\gamma}\right)\ev{\sigma_n^y}+\Delta_n^{in}\ev{\sigma_n^x}\notag\\
                          &-\frac{\Omega}{2\gamma}\sum_m J_{nm}\ev{\sigma_n^x\sigma_m^y} \label{eq:sigmaySimp}
\end{align}
thus leading to (\ref{eq:EOMQuantum}). In (\ref{eq:sigmaySimp})-(\ref{eq:sigmaxSimp}), the time dependence is implied. \par 
Let us now define $s_n(t)\equiv \ev{\sigma^y_n(t)}$ and $I_n(t) \equiv \ev{\sigma^x_n(t)}$ so that $\bs{s}(t)=(s_1(t),...,s_N(t))$ and $\bs{I}(t)=(I_1(t),...,I_N(t))$.
We see that (\ref{eq:sigmaySimp})-(\ref{eq:sigmaxSimp}) match (\ref{eq:I})-(\ref{eq:r}) where 
\begin{align}
    G_n &= -\Delta_n s_n - \frac{\Omega}{2\gamma}\sum_{m}J_{nm}\ev{I_n I_m},    && \tau_I^{-1} = \frac{\gamma}{2},\\
    F_n &= \Delta_n I_n - \frac{\Omega}{2\gamma}\sum_{m}J_{nm}\ev{I_n s_m},  && \tau_s^{-1} = \frac{\gamma}{2}+\frac{\Omega^2}{4\gamma}.
\end{align}
Equations (\ref{eq:sigmaxSimp})-(\ref{eq:sigmaySimp}) allow us to naturally interpret $\ev{\sigma_n^y}$ as the firing rate of the $n^{th}$ neuron, and $\ev{\sigma_n^x}$ as the current. That is, the rate of the pre-synaptic neuron $\ev{\sigma^y_k}$ amounts to a current in the post-synaptic neuron $\ev{\sigma^x_n}$ that drives its rate $\ev{\sigma_n^y}$. \par 
Equations (\ref{eq:sigmaxSimp})-(\ref{eq:sigmaySimp}) comprise a system of coupled quadratic differential equations, where the quadratic terms arise from the nontrivial commutation relation of the Pauli-operators $[\sigma_n^\alpha,\sigma_m^\beta]=i\delta_{\alpha\beta}\epsilon_{\alpha\beta\gamma}\sigma_n^\gamma$ where $\epsilon_{\alpha\beta\gamma}$ is the Levi-Civita symbol. These quadratic terms in (\ref{eq:sigmaxSimp})-(\ref{eq:sigmaySimp}) make a qRNN a powerful computational system similar to how the functions $G$ and $F$ make an RNN a powerful computational system.

\section{Memory and quantum many-body scars}\label{A:QScars}
\begin{figure}
\centering
\includegraphics[width=\linewidth]{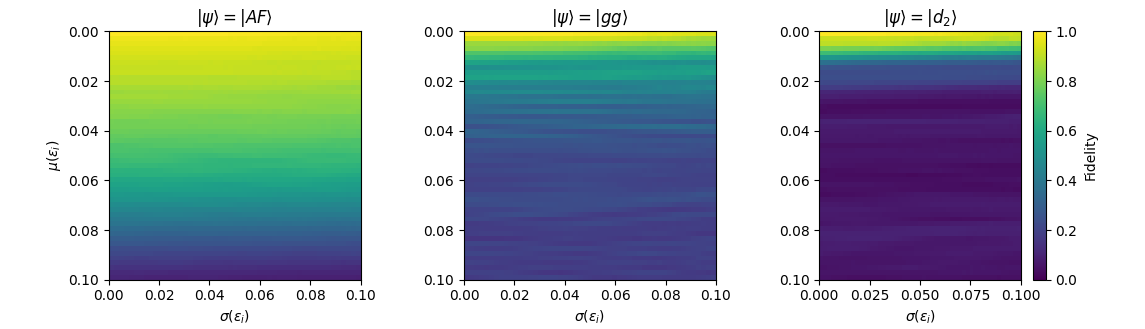}
\caption{Fidelities with the initial state after evolving for $n=100$ cycles of noisy, kicked dynamics. The fidelity is defined as $F\equiv |\langle \psi|\psi(2n\tau)\rangle|^2$ where $|\psi\rangle$ is the initial state. Here, we used $L=8$ Rydberg atoms and define $|AF\rangle = |grgrgr\rangle$, $|gg\rangle=|gggggggg\rangle$, and $|d_2\rangle=|grggggrg\rangle$. The Neel state $|AF\rangle$ is robust to the noise in the drive since this state is invariant to decoherence up to second order in $\epsilon_i$.}
\label{fig:Fidels}
\end{figure}

\renewcommand{\theequation}{C.\arabic{equation}}
As described in the main text, and more thoroughly discussed in Ref. \cite{maskara2021discrete}, the scaring behavior of the kicked PXP-model is robust to fixed imperfections in the drive. The robustness persist even for random noise. Fig. \ref{fig:Fidels} exemplifies the overlap with the initial condition for a noisy, kicked PXP-model for different values of $\epsilon$ and $\sigma_{in}$, which is a natural extension of the model in \cite{maskara2021discrete}. The Neel state $|AF\rangle$ exhibits robust revivals invariant of $\sigma^2_{in}$. This fact can be explained with the effective theory presented below.

To understand the robustness of the quantum scaring behavior in the Rydberg reservoir it is instructive to seek an effective description of the system's evolution. Recall that a cycle is defined as two imperfect applications of $\chi_\tau$. The Hamiltonian in (\ref{eq:PXP1}) produces the single-cycle unitary 
\begin{equation}\label{eq:U1}
    U_\tau(\epsilon_1,\epsilon_2) = e^{-i\epsilon_2 \hat{N}}\chi_\tau e^{-i\epsilon_1 \hat{N}}\chi_\tau = e^{-i\epsilon_2\hat{N}}e^{-i\epsilon_1\chi_\tau \hat{N}\chi_\tau}
\end{equation}
where we use the fact that $\chi_\tau$ is both Hermitian and unitary. Using the Baker-Campbell-Hausdorf formula to second order in $\epsilon_i$, we can rewrite (\ref{eq:U1}) as
\begin{equation}
     U_\tau(\epsilon_1,\epsilon_2) \approx e^{-i(\epsilon_2 \hat{N} +\epsilon_1 \chi_\tau \hat{N}\chi_\tau)}.
\end{equation}
A state $\rho(n)$ evolves to $\rho(n+1) = U_\tau(\epsilon_1,\epsilon_2)\rho(n)U^\dagger_\tau(\epsilon_1,\epsilon_2)$ after a cycle. Expanding this to second order in $\epsilon_k$ and using the fact that $\langle \epsilon_k\rangle= \epsilon$ and $\langle \epsilon_k\epsilon_l\rangle = \sigma_{in}^2\delta_{kl}$, we obtain the average evolution of the state
\begin{align}\label{eq:L1}
    \rho(n+1)-\rho(n)&= -i\varepsilon[H^{+}, \rho(n)] \notag \\
    &+\sigma_{in}^2\left(\hat{N}\rho(n)\hat{N} - \frac{1}{2}\{\hat{N}^2,\rho(n)\}\right) \notag\\
    &+\sigma_{in}^2\left(\chi_\tau \hat{N}\chi_\tau\rho(n) \chi_\tau \hat{N}\chi_\tau\right.\notag\\
    &\left.- \frac{1}{2}\{\chi_\tau \hat{N}^2\chi_\tau,\rho(n)\}\right).
\end{align}
Here, $\{A,B\} = AB+BA$ denote commutators and anti-commutators respectively. We define $H^+ = \hat{N}+\chi_\tau \hat{N}\chi_\tau$. For times $T\gg 2\tau$, we can take (\ref{eq:L1}) to be a Lindbladian evolution since the noise satisfies the Markovian properties. We can rewrite (\ref{eq:L1}) as 
\begin{align}
    \dot{\rho} &= \mathcal{L}_{\epsilon,\sigma}(\rho) \\
    \mathcal{L}_{\epsilon,\sigma}(\cdot) &= -i\frac{\varepsilon}{2\tau}[H^{+}, \cdot]+\frac{\sigma_{in}^2}{2\tau}D^+(\cdot)+\frac{\sigma_{in}^2}{2\tau}D^-(\cdot)\\
    D^{\pm}(\cdot) &= H^{\pm}\cdot H^{\pm}+\frac{1}{2}\{H^{\pm}H^{\pm},\cdot\}
\end{align}
where $H^{-} = \hat{N}-\chi_\tau \hat{N}\chi_\tau$. For $\tau = 1.51\pi$, the Neel states are approximately simultaneous eigenstates of $\chi_\tau \hat{N} \chi_\tau$ and $\hat{N}$ with eigenvalues $N$ for a system of size $N$. Thus, they are simultaneous eigenstates of $H^{\pm}$ and so 
\begin{equation}\label{eq:AFZero}
    \mathcal{L}_{\epsilon,\sigma}(|AF\rangle\langle AF|) \approx 0, \quad \mathcal{L}_{\epsilon,\sigma}(|AF'\rangle\langle AF'|) \approx 0.
\end{equation}
Therefore, the Neel states are steady states. It is worth noting that $\mathcal{L}_{\epsilon, \sigma}$ captures the pre-thermal evolution. Ultimately, higher-order effects in $\epsilon_k$ takeover and lead to the thermalization of the Neel states similar to the results in \cite{maskara2021discrete} and as seen in Fig. \ref{fig:Memory}. Nonetheless, the thermalization of the Neel states is delayed relative to other states due to (\ref{eq:AFZero}). \par  

\begin{figure}
\centering
\includegraphics[width=0.8\linewidth]{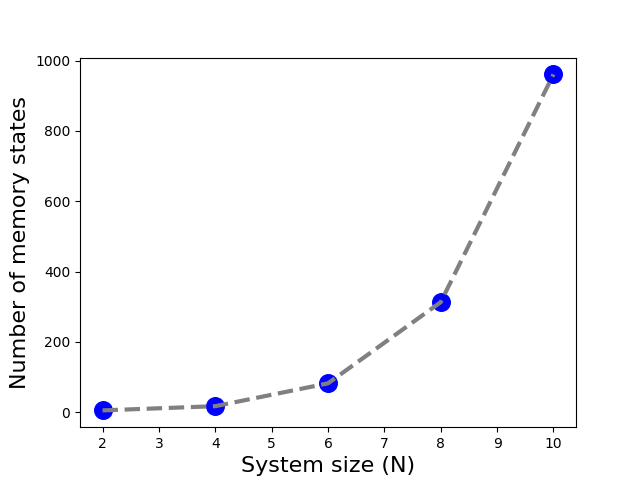}
\caption{Number of zero eigenvalues of the super-operator $\mathcal{L}_{\epsilon,\sigma}$ as a function of the system size. $\mathcal{L}_{\epsilon,\sigma}$ describes the effective dynamics of a Rydberg reservoir composed of kicked Rydberg atoms. The number of zeros surpasses the linear number of memories available in the Hopfield network.}
\label{fig:zeros}
\end{figure}
\begin{figure}
\centering
\includegraphics[width=\linewidth]{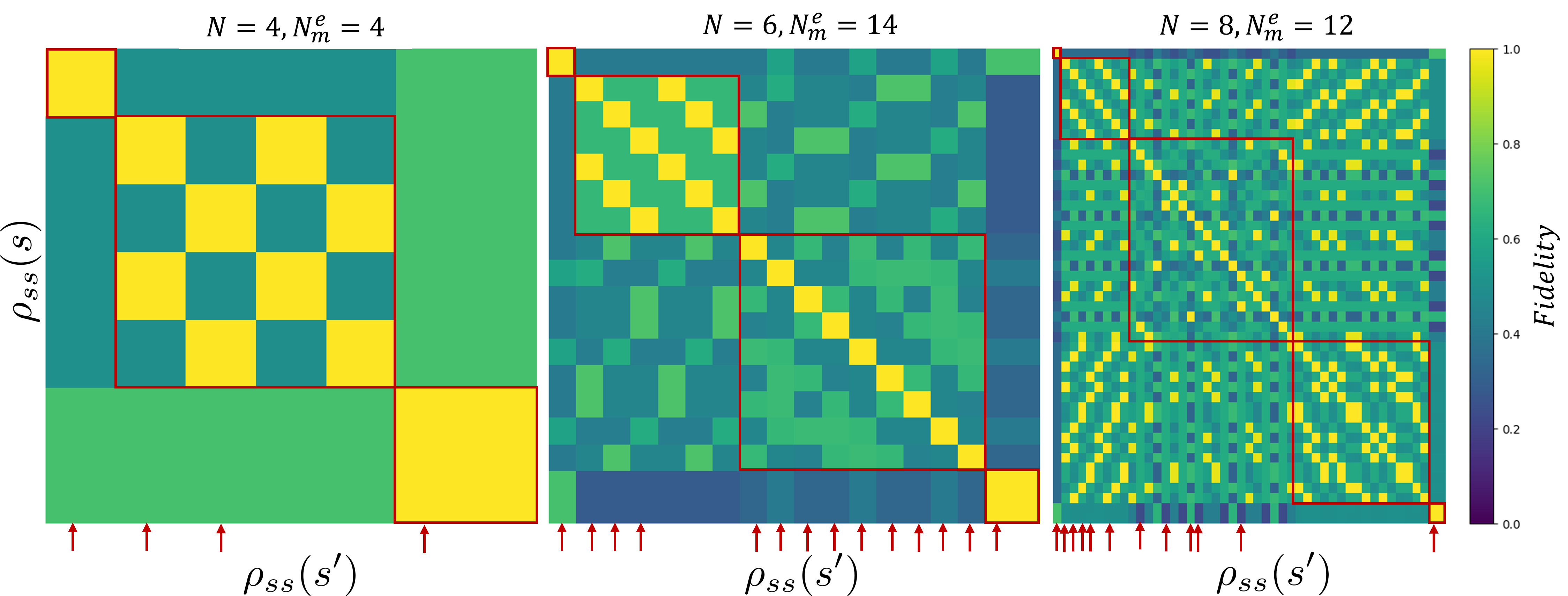}
\caption{Empirical memory states $\rho_{ss}(s)$ obtained from evolving the initial states $|s\rangle$ which are basis states of the Rydberg blockaded Hilbert space. $N_m^e$ denotes the number of memories found using this procedure. The different plots show the fidelities $F(\rho_{ss}(s), \rho_{ss}(s'))$ between different steady states. The red squares delimit the basis states with different number of Rydberg excitations starting with the zero excitation sector on the top-left square and ending with $N/2$ excitations sector on the bottom-right square. Red arrows denote initial configurations for each of the $N^{e}_m$ memories found empirically. While this procedure produces a number of memory states smaller than the number of zeros of $\mathcal{L}_{\epsilon,\sigma}$, $N_{m}^e>N$, a bound unattainable by common classical RNNs. }
\label{fig:steadystates}
\end{figure}
Moreover, any density matrix $\rho_{ss}$ in the kernel of $\mathcal{L}_{\epsilon, \sigma}$ can be used as a memory state. Expressing $\mathcal{L}_{\epsilon, \sigma}$ as a super-operator on density matrices, we can look at its spectrum which is in general complex. Fig. \ref{fig:zeros} shows the number of zero eigenvalues of $\mathcal{L}_{\epsilon, \sigma}$ for different system sizes $N$. The number of zeros scales larger than linearly on $N$. Therefore, a quantum reservoir evolving under $\mathcal{L}_{\epsilon,\sigma}$ may have a larger number of memory states than a classical RNN. To prepare these states, we propose to initialize the reservoir on different string configurations $|s\rangle$ satisfying the Rydberg blockade constraint. For example, one can have $s=rgg..g$ while $s=rrg...g$ is not allowed. The system is left to evolve for some time $T_{ss}$ to reach a steady state $\rho_{ss}(s)$ which can then be used as memory. Different initial strings can lead to different steady states as exemplified in Fig. \ref{fig:steadystates}. Fig. \ref{fig:steadystates} shows the fidelity between $\rho_{ss}(s)$ and $\rho_{ss}(s')$ defined by the trace norm
\begin{equation}
    F(\rho_{ss}(s), \rho_{ss}(s')) = \left( \text{Tr}\sqrt{\sqrt{\rho_{ss}(s)}\rho_{ss}(s')\sqrt{\rho_{ss}(s')}}\right)^2.
\end{equation}
The red arrows in Fig. \ref{fig:steadystates} indicate the different memory states obtained by this scheme. It's worth noting that this scheme offers us an empirical number of memories $N^e_m$ that scales at most as $\phi^N$, where $\phi\approx 1.62$ is the Golden ratio since that's the number of basis states respecting the Rydberg blockade. We see that $N^e_m>N$ in all instances, a bound unattainable by classical RNNs such as the Hopfield network \cite{folli2017maximum}. However, this scheme relies on an efficient way to recognize the different memory states through measurements, a question that we leave for future investigations.

\section{Experimental values, and numerical simulations}\label{A:Numerics}
\begin{figure}
    \centering
    \includegraphics[width=0.8\linewidth]{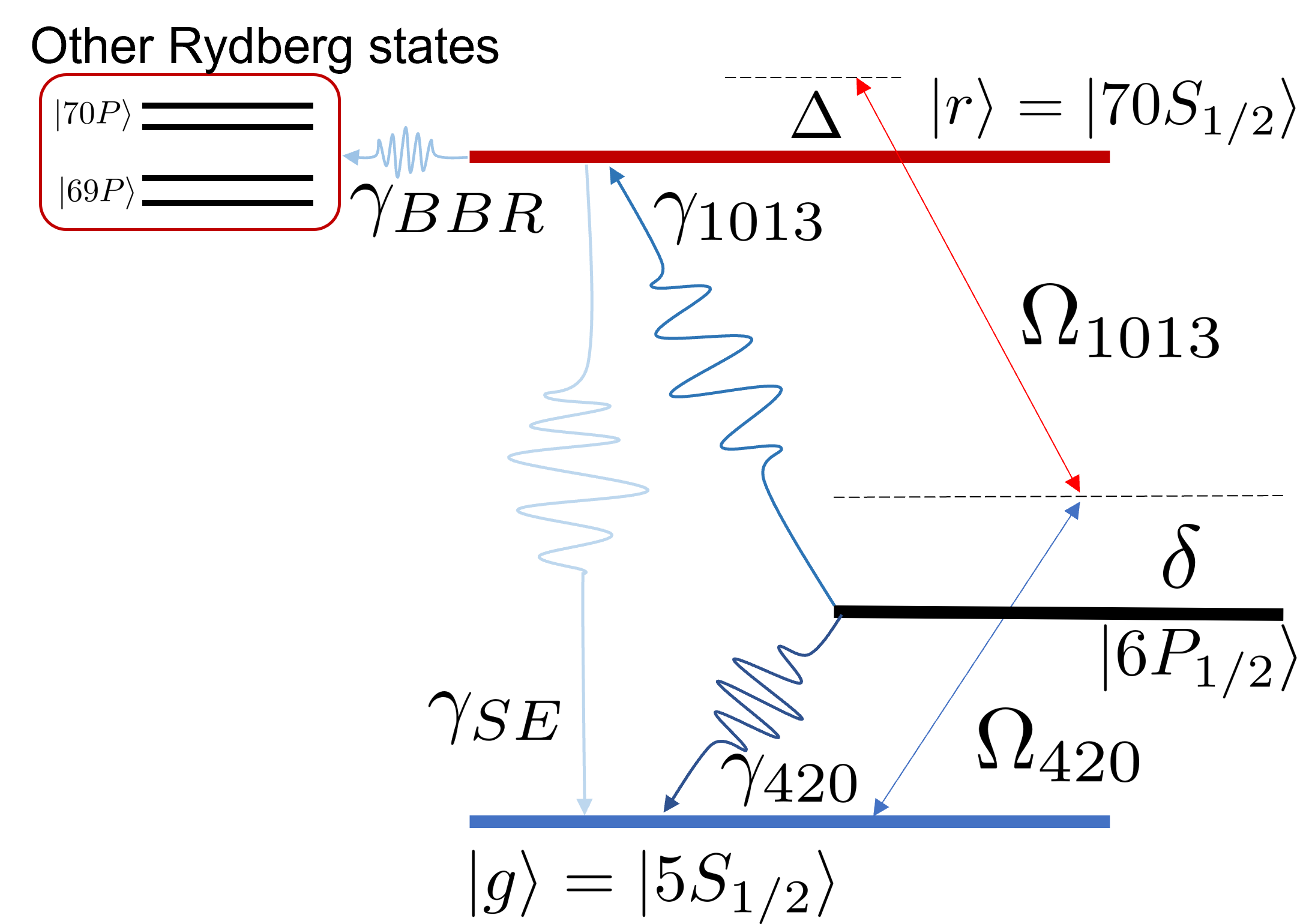}
    \caption{\edit{Schematic of Rydberg atoms as used in Ref. \cite{Ebadi2020}. The ground state $|g\rangle=|5S_{1/2}\rangle$, and the Rydberg state $|r\rangle = |50S_{1/2}\rangle$ are coupled via a two-photon transition. An off-resonance 420 nm laser ($\Omega_{420}=2\pi\times160\text{ MHZ}$, $\delta= 2\pi\times 1 \text{ GHz}$) couples $|g\rangle$ with the intermediate $|6P_{3/2}\rangle$ state, and a 1013 nm laser ($\Omega_{1013}=2\pi\times50\text{ MHz}$ couples the intermediate state and $|r\rangle$ creating an effective drive between $|g\rangle$ and $|r\rangle$ at rate $\Omega = \frac{\Omega_{420}\Omega_{1013}}{\delta}=2\pi\times4.2 \text{MHz}$. Four spontaneous emission processes are at play: emission to nearby Rydberg atoms due to black-body-radiation at a rate $\gamma_{BBR}=2\pi/(250 \text{ }\mu\text{s})$, photon-scattering out of the intermediate state into the ground state at rate $\gamma_{420}=2\pi/(20 \text{ }\mu\text{s})$ and into the Rydberg state at rate $\gamma_{1013}=2\pi/(150 \text{ }\mu\text{s})$, and spontaneous emission from $|r\rangle$ to $|g\rangle$ at rate $\gamma_{SE}=2\pi/(375 \text{ }\mu\text{s})$. 
    Since $\gamma_{BBR}+\gamma_{SE}+\gamma_{1013}= 2\pi/(75 \text{ }\mu\text{s})$ is smaller than $\gamma_{420}$, the leading source of decoherence for short periods of time ($< 10 \text{ }\mu\text{s}$) is due to the  $\gamma_{420}$ decay.}}
    \label{fig:Rydberg_atom}
\end{figure}
\edit{
In this section, we outline the details of the experimental values used for the numerical simulation of Sec. \ref{sec:BioTasks}. Firstly, for simulating Rydberg atoms we use the experimental values in Ref. \cite{Ebadi2020} for concreteness ( see Fig. \ref{fig:Rydberg_atom}). In this experimental platform, a two-photon transition couples $|g\rangle=|5S_{1/2}\rangle$ and $|r\rangle=|50S_{1/2}\rangle$ via an off-resonance state $|6P_{3/2}\rangle$. For this setup, and for short periods of simulation ($< 10 \text{ }\mu\text{s}$), the dominant source of decoherence is photon-scattering processes out of the intermediate state. Using the fact that the intermediate state is off-resonance, we can adiabatically eliminate it to produce an effective decay operator (see Sec. IV.B in Ref. \cite{reiter2012effective}) 
\begin{equation}
    \sigma^{-}_{eff} = \frac{\sqrt{\gamma_{420}}}{2\delta}|g\rangle\left(\Omega_{420}\langle g|+\Omega_{1013}\langle r|\right)
\end{equation}
which is an effective spontaneous emission from $|r\rangle$ to $|g\rangle$ accompanied by decoherence on the ground state. \par 
We chose $\Omega = 4.2 \text{ MHz}$. Additionally, a pair of $|r\rangle$ atoms interact with a strength $C_{6} = 862.9$ GHz$(\mu\text{m})^6$. We used the \texttt{PairInteraction} python package from \cite{Weber2017} to determine that a pair of $|r\rangle=|70S_{1/2}\rangle$ and $|r'\rangle=|73S_{1/2}\rangle$ has a similar interaction strength of $C_6^{rr'} = -836.6 \text{ GHz}(\mu\text{m})^6\approx -C_6$. We used this interaction to model the inhibitory and excitatory neurons in Sec. \ref{sec:multitasking} ($V_{n_Q, n_Q} = V$, $V_{n_Q, n_Q'}=-V$). We denote $V = C_6/a^6_0$ where $a_0$ is tuned to give us different nearest neighbor interaction strengths.\par 
Next, we explain and report the numerical parameters chosen for each of the biological tasks. 

\subsection{Multitasking}
Our scheme to encode inhibitory and excitatory neurons relies on approximating (13), and as a result, one needs the “inhibitory neurons” to be as far away as possible from each other such that they do not interact positively with each other. For this reason, this task uses a 1D open chain of atoms separated by a distance $a_0$ with the inhibitory neurons being at opposite ends of the chain and in the bulk with maximum spacing from each other. The input neurons are chosen to be the two at one end of the chain, while the output neuron is chosen to be at the opposite end of the chain. This choice was made to ensure that the input neurons interact with the whole chain before readout.\par 
The inputs are uniformly sampled from $\{0,2\pi\}$ MHz with added Gaussian noise $\sigma_{in}=0.1$, and all $\Delta t$ sampled from a Gaussian with average $\langle \Delta t\rangle \in [0,5]$ $(\mu$s) and standard deviation $\sigma_{in}$. For each size of the network and number of inhibitory neurons, We choose $a_0$ such that the separation between inhibitory neurons $d_{max}$ results in $V/d_{max}^2=10^{-2}$. For example, for the case of 4 neurons and two inhibitory neurons on either end, note that one needs $V/3^6 = 10^{-2}$, which amounts to choosing $V =  7.2$ MHz. Note that this value of $V$ is of the order of magnitude of $\Omega=4.2$ MHz and so the reservoir, in this case, is well into the non-classical regime. \par 
The learned parameters in the output linear map $W^\out$ which in this case is a matrix in $\mathbb{R}^{3+1\times 1}$ with the last row representing a bias term. Note that the dimension of the map is so because only one neuron is measured but three functions have to be fitted. 

\subsection{Decision making}
In classical RNNs tasks such as decision making and working memory require connectivity between all neurons. Since our connectivity is limited by physical constraints, an open 2D square lattice structure was chosen to prevent neurons from being isolated from the rest. Moreover, a 2D square lattice is experimentally friendly. In our case, we use an open $2\times3$ lattice with the two input neurons being at the top-left corner of the chain, and two output neurons being at the bottom-right corner. Again, this architecture was chosen so that the input neurons have to interact with the rest of the system before readout. We use $V=2\pi\times 10$ MHz for our simulations, and choose $\Delta t= 2\pi/V$ as the time the inputs are turned on as that's the timescale in which the input atoms entangle with the rest of the chain.\par 
The inputs are uniformly sampled from $\{0, \pi/2, \pi, 3\pi/2, 2\pi\}$ (MHz) with added Gaussian noise $\sigma_{in}=0.1$. In this task, the time that the stimuli are turned on $\Delta t$ is fixed to a mean of $\langle \Delta t \rangle = 0.1$ $\mu$s and with added Gaussian noise $\sigma_{in}=0.1$. In this task, we optimize over the linear output map $W^\out$, a matrix in $\mathbb{R}^{1+1,2}$ since one function is fitted and two neurons are measured. Additionally, we train the output time $t_{out}$ after the stimuli are turned off and before the network is probed to come up with an input that is satisfied (\ref{eq:ytarg_decision}). To do the optimization, we make use of the Nelder-Mead algorithm \cite{nelder1965simplex}. \par 
In order to compute the psychometric response plotted in Fig. \ref{fig:decision making}B, we measure the expectation values on the two output neurons and produce the vector $\bs{r}(\Delta^{in}_1,\Delta^{in}_2)= (\langle \sigma_{out1}^y \rangle,\langle \sigma_{out2}^y\rangle,1)$ which depends on the inputs $\Delta^{in}_{1,2}$, as well as the temporal parameters $(\Delta t, t_{out})$. We then compute $y^{\out}(\Delta^{in}_{1,2})=W^\out\cdot \bs{r}(\Delta^{in}_{1,2})$ and $(W^\out,t_{out})$ are optimized such that $y^{\out}(\Delta^{in}_{1,2})\approx y^{targ}$ in \ref{eq:ytarg_decision}. The optimization is done by generating about 40,000 different values of $\Delta_{1,2}^{in}$ of different levels of contrast $|\Delta^{in}_1-\Delta^{in}_2|$ ranging from 0 to 1 MHz. Once the optimization is done, we look at the loss towards $\Delta_2^{in}$, which is obtained as the error in classifying $\Delta_2^{in}$ as greater than $\Delta_1^{in}$ when indeed $\Delta_2^{in}>\Delta_1^{in}$. The error is quantified using the mean-square loss in (4).

\subsection{Working memory}
This task's setup is identical to the decision-making task except that the two inputs are separated by a delay time $t_{delay}$. The values of the interaction strength $V$ used for Fig. \ref{fig:working-memory} are $V=2\pi \times 10 $ MHz and $V=2\pi \times 0.1 $ MHz corresponding to $V/\Omega>1$ and $V/\Omega<1$ respectively. The former of which sets us in the Rydberg blockaded regime while the latter is not. In this task, the times $\Delta t$ and $t_{delay}$ are fixed up to an added Gaussian with noise $\sigma_{in}=0.1$. In this task, we optimize over the linear output map $W^\out$, a matrix in $\mathbb{R}^{1+1,2}$ since one function is fitted and two neurons are measured.\par 

\subsection{Long-term memory}
Although quantum scars are known to exist in other geometries and dimensions \cite{serbyn2021quantum}, for this task we use a 1D chain of Rydberg atoms since for this case quantum many-body scars have been experimentally observed \cite{Bernien2017, bluvstein2021controlling}. Furthermore, our chain has periodic boundary conditions to avoid edge effects. Since we know that scars are robust to decoherence, we set $\gamma_{420}=0$ so that we can evolve our states for longer periods. The number of cycles $n$ in Fig. \ref{fig:Memory} corresponds to $n$ evolutions under the PXP Hamiltonian for a time $2\tau=1.51\times \pi \Omega^{-1}$. In this case, we take $V\gg \Omega$ and renormalized $\Omega=1$. The noisy field in (\ref{eq:PXP1}) is sampled according to $\epsilon_k\sim N(\mu=0.1, \sigma=0.1)$. The input $m$ is sampled as a fair random coin. Lastly, after each number of cycles $n$, the only trained parameter is $W_n^\out\in \mathbb{R}^{1+1\times 1}$ since only one atom is probed to calculate an answer as to the input $m$. \par
}

\end{document}